\documentclass[%
 reprint,
 superscriptaddress,
 amsmath,amssymb,
 aps,
 prd,
floatfix,
]{revtex4-2}

\usepackage{graphicx}
\usepackage{epsfig}
\usepackage{appendix}
\usepackage{dcolumn}
\usepackage{bm}
\usepackage{isotope}
\usepackage{color}
\usepackage{graphicx,epsfig}
\usepackage{xcolor}
\usepackage{textcomp}
\usepackage[nowatermark]{fixmetodonotes}

\begin{document}

\title{Quasielastic Electromagnetic Scattering Cross Sections and World Data Comparisons in the {\fontfamily{qcr}\selectfont GENIE} Monte Carlo Event Generator}

\author{J. L. Barrow}
\email{jbarrow3@vols.utk.edu}
\affiliation{Department of Physics and Astronomy, The University of Tennessee, Knoxville, TN  37996,  USA}
\affiliation{Fermi National Accelerator Laboratory, Batavia, IL 60510, USA}
\author{S. Gardiner}
\affiliation{Fermi National Accelerator Laboratory, Batavia, IL 60510, USA}
\author{S. Pastore}
\affiliation{Department of Physics, Washington University in Saint Louis, Saint Louis, MO 63130}
\affiliation{McDonnell Center for the Space Sciences at Washington University in St. Louis, MO, 63130, USA}
\author{M. Betancourt}
\affiliation{Fermi National Accelerator Laboratory, Batavia, IL 60510, USA}
\author{J. Carlson}
\affiliation{Los Alamos National Laboratory, Los Alamos, NM, 87545 USA}

\date{\today}

\begin{abstract}
The usage of Monte Carlo neutrino event generators (MC$\nu$EGs) is a norm within the high-energy $\nu$ scattering community. The relevance of quasielastic (QE) energy regimes to $\nu$ oscillation experiments implies that accurate calculations of $\nu A$ cross sections in this regime will be a key contributor to reducing the systematic uncertainties affecting the extraction of oscillation parameters. In spite of this, many MC$\nu$EGs utilize highly phenomenological, parameterized models of QE scattering cross sections. Moreover, a culture of validation of MC$\nu$EGs against prolific electron ($e$) scattering data has been historically lacking. In this work, we implement new $e A$ cross sections obtained from nuclear ab initio approaches in GENIE, the primary MC$\nu$EG utilized by the FNAL community. In particular, we utilize results from Quantum MC methods which solve the many-body nuclear problem in the Short-Time Approximation (STA), allowing consistent retention of two-nucleon dynamics which are crucial to explain available nuclear electromagnetic (electroweak) data over a wide range of energy and momentum transfers. This new implementation in GENIE is fully tested against the world QE electromagnetic data, finding agreement with available data below $\sim2\,$GeV of beam energy with the aid of a scaling function formalism. The STA is currently limited to study $A\leq12$ nuclei, however, its semi-inclusive multibody identity components are exportable to other many-body computational techniques such as Auxiliary Field Diffusion MC which can reach $A\leq40$ systems while continuing to realize the factorization contained within the STA's multinucleon dynamics. Together, these developments promise to make future experiments such as DUNE more accurate in their assessment of MC$\nu$EG systematics, $\nu$ properties, and potentially empower the discovery of physics beyond the Standard Model.
\end{abstract}

\maketitle

\section{Introduction}
\subsection{Future $\nu$ Oscillation Study Requirements}

There is great importance for the particle physics community in the difficult task of mapping experimentally observed final state neutrino ($\nu$) properties and energies onto initial $\nu$ states given the presence of oscillations and the complexity of chosen target nuclear systems~\citep{Alvarez-Ruso:2017oui}. Indeed, whether these $\nu$’s originate in a beam or the atmosphere, any lack of capacity in the reconstruction of these quantities can lead to misinterpretations of the true physics of the system under study, potentially distorting future results of the global short- and long-baseline $\nu$ oscillation program. Many technicalities and their interrelations limit the interpretive certainty of any experiment’s results, including the $\nu$ cross section model and its dependence on the assumed structure of the nuclear target with (or without) the inclusion of \textit{multinucleon correlations and interference effects}, the $\nu$ flux and beam divergence model, the intranuclear cascade (final state interactions) model, and the detector’s capability (response) in efficiently reconstructing the topology of a $\nu$ event at particular kinematics. Many of these can currently only be efficiently simulated using Monte Carlo $\nu$ event generators (MC$\nu$EGs), a popular candidate being {\fontfamily{qcr}\selectfont GENIE}~\citep{Andreopoulos:2009rq}, and experiments rely on these and other types of computation to simulate their $\nu$ beam, $\nu$ interactions, intranuclear cascade, and observable final state topologies given modeled detector responses. All of these tools are necessary components for precise measurements of $\nu$ properties such as $CP$-violation ($\delta_{CP}$) and the ordering of the $\nu$ masses.

In this work, we focus on one of the above-mentioned components: cross sections with complex nuclear structure and multinucleon interactions \textit{intact}. To outline this work in brief:
\begin{enumerate}
    \item We briefly explain the {\it ab initio} methods used within the quantum Monte Carlo Short-Time Approximation (QMC STA)~\citep{Pastore:2019urn} and how these can be applied in the calculation of electromagnetic nuclear response functions and response densities for light nuclei;
    \item We discuss our new, holistic framework within {\fontfamily{qcr}\selectfont GENIE} based on calculated electromagnetic (electroweak) nuclear response functions and supplemented by interpolation schemes to compute double differential electron scattering cross sections;
    \item We show comparisons of these responses and double differential cross sections against abundant electromagnetic scattering data to assess the validity of both the {\fontfamily{qcr}\selectfont GENIE} implementation and the theoretical nuclear response function inputs. Similar studies have recently been performed for existing \mbox{\fontfamily{qcr}\selectfont GENIE} cross section models by multiple groups \cite{PhysRevD.102.053001,Papadopolou2020}.
\end{enumerate}
This is a foundational work where we test our framework and verify that the events generated by {\fontfamily{qcr}\selectfont GENIE} are fully consistent with the inputs provided by the underlying theoretical calculations of nuclear responses from the STA. More broadly, this work will construct a solid basis for future implementations of $\nu$-nucleus responses in the MC$\nu$EG. When using a consistent microscopic model of $V$ and $V-A$ lepton-nucleus interactions, these will allow scientists to better estimate the precision of $\nu$ scattering event samples produced by MC$\nu$EGs which in-turn are used to understand experimental $\nu$ cross sections and oscillation parameters; such increases in the precision of these measurements may permit the necessary resolution to discover physics beyond the Standard Model (BSM).

\subsection{Quasielastic Scattering Overview}
Quasielastic (QE) scattering, or when a particle probes a nucleus by transferring energy and momentum {\it primarily} to a single nucleon, is a key interaction process observed at both current electron-scattering facilities, {\it e.g.}, Jefferson Laboratory~\citep{JLAB},  as well as current and future short- and long-baseline $\nu$ oscillation experiments~\citep{mb_web,nova_web,DUNE}. However, the majority of the models utilized by MC event generators in this energy regime are generally highly phenomenological. Typically, an effective single-nucleon cross section is implemented which inherently ignores important high Bjorken-$x$ interactions visible as missing energy or momentum via two-nucleon short-range correlations~\citep{Subedi:2008zz,Fomin:2017ydn}. This implies that a large portion of the truly quantum behavior at play within the nucleus being probed is partially or entirely ignored, including interference terms and tensor forces which mediate two-body dynamics and can create observable two-nucleon topologies in detectors, independent of final state interactions (FSIs). Overlooking these important dynamical components can lead to a suppression of the cross sections, which can in turn make experimental measurements \textit{appear} enhanced in strength, perhaps leading accidentally to interpretations of extraordinary physics.

The QMC STA~\citep{Pastore:2019urn}, adopted in the present work, incorporates these nontrivial multinucleon dynamics directly within electromagnetic nuclear response~\textit{densities} and associated nuclear response functions. The latter are given as functions of the energy, $\omega$, and three-momentum transfer, $|{\bf q}|$. Using precomputed tables of these responses, one can interpolate across $(\omega, |\mathbf{q}|)$ space to calculate inclusive double differential and total QE cross sections where effects from two-body physics and enhancements can be observed. Since a formalism involving these nuclear response functions is common to many models of lepton-nucleus scattering, a software framework which takes them as input and uses them to produce simulated events allows competing models to be compared easily within a MC$\nu$EG.  Given the complexity of the codes generally utilized to solve the many-body nuclear problem, direct implementation of the most realistic calculations in a MC$\nu$EG is impractical. Tables of precomputed nuclear responses allow for efficient event generation while preserving the physics content of sophisticated inclusive cross section models~\cite{Barrow:2020gzb}. Though this work is focused on electromagnetic scattering on very light nuclei, QMC STA methods are \textit{directly} extendable to include up to $A\leq12$ nuclei for electromagnetic and electroweak scattering; other known QMC computational methods, such as Auxiliary Field Diffusion Monte Carlo~\citep{Carlson:2014vla}, can similarly maintain the interference and two-body contributions at play within the QE cross section, while being exportable to the $A\leq40$ systems most important for future experimental programs such as the Deep Underground Neutrino Experiment (DUNE)~\citep{DUNE}. This further motivates the creation of a universal input framework for use by theorists to more easily incorporate their work into experimental MC event production and analysis chains.

As a \textit{start} to this long-term computation, simulation, and validation program outside and within the {\fontfamily{qcr}\selectfont GENIE} collaboration, here we consider inclusive QE scattering of electrons on
\isotope[4][2]{He}, and validate the behavior of the QMC STA within the {\fontfamily{qcr}\selectfont GENIE} MC event generator across the QE-regime on publicly maintained world inclusive QE electron scattering data~\citep{Benhar:2006wy,EMQEOnlineDataSet}. Further, we offer some predictions of $nn$, $pp$, and $np$ contributions to the cross sections, which we hope to be useful for current and future electron scattering experiments, while also hinting a path forward for the $\nu$ community. In closing, we emphasize that the main point of this work is to validate our framework. This will set a solid basis for future developments in the {\fontfamily{qcr}\selectfont GENIE} MC$\nu$EG.

\subsection{{\fontfamily{qcr}\selectfont GENIE} Overview}

{\fontfamily{qcr}\selectfont GENIE} (Generates Events for Neutrino Interaction Experiments) \citep{Andreopoulos:2009rq} is a collaboratively written and maintained suite of MC event generator and model tuning \citep{Andreopoulos:2015wxa} packages used by many $\nu$ experiments, including \mbox{MINERvA}~\citep{minerva_web}, MicroBooNE~\citep{mb_web}, the Short-Baseline Near Detector~\citep{sbnd_web}, and DUNE \citep{DUNE,Abi:2020wmh,Abi:2020evt,Abi:2020loh}. Within \mbox{\fontfamily{qcr}\selectfont GENIE}, lepton-nucleus interactions are modeled as a two step process using the impulse approximation; interactions occur on individual bound and moving nucleons, and outgoing hadrons resulting from the primary interactions propagate through the nucleus and are subject to FSIs. {\fontfamily{qcr}\selectfont GENIE} is an event generator which seeks to provide comprehensive modeling for all nuclear targets and leptons of all flavors from MeV to PeV energy scales \citep{Andreopoulos:2009rq}. Using C++, XML, and CERN ROOT~\citep{Antcheva:2011zz}, {\fontfamily{qcr}\selectfont GENIE} offers modularity in its configurations and code design, and the collaboration encourages scientists to contribute new model implementations using their platform in the form of ``Incubators". The work described in this article springs from just such an Incubator.

As a means of benchmarking $\nu$ cross section models against electron scattering data, the {\fontfamily{qcr}\selectfont GENIE} interface for consuming nuclear response tables was recently generalized to handle generation of both neutrino and electron scattering events on an equal footing.

\subsection{\label{sec:QEInclusiveXSec}QE Inclusive Cross Sections}

The QE inclusive-scattering cross section of electrons and $\nu$s on nuclei can be considered in terms of nuclear electromagnetic or electroweak response functions. Under the assumption that the lepton-nucleus interaction is dominated by the exchange of a single virtual photon which couples to the nucleus' electromagnetic charge and current, the  electron-scattering cross section of interest in this work is given by~\citep{deforest66,Carlson:1997qn,Carlson:2001mp,Bacca:2014tla}
\begin{equation}
\label{eq:xsec}
    \frac{d^2\,\sigma}{d\,\omega d\,\Omega}=\sigma_M\,\left[v_L\,R_L({\bf q},\omega)+v_T\,R_T({\bf q},\omega) \right]\, ,
\end{equation}
where $\omega$ and ${\bf q}$ are  the  energy and three-momentum transfer, respectively, and $\sigma_M$ is the Mott cross section defined as:
\begin{equation}
\label{eq:Mottxsec}
    \sigma_M=\left( \frac{\alpha\, {\rm cos}{\theta/2}}{2\, \epsilon_i \, {\rm sin}^2{\theta/2}} \right)^2 \,.
\end{equation}
In Eq.~(\ref{eq:Mottxsec}), $\alpha$ is the fine structure constant, $\theta$ the electron scattering angle, and $\epsilon_i$ the initial electron energy. The lepton's kinematic factors are defined as
\begin{equation}
    v_L=\frac{Q^4}{q^4}\, , \quad \quad \quad v_T=\frac{Q^2}{2\,q^2}\,+{\rm tan}^2\frac{\theta}{2} ,
\end{equation}
where $Q$ is the four-momentum transfer. The two nuclear electromagnetic response functions, namely the longitudinal and the
transverse, are  schematically given by
\begin{eqnarray}
       R_\alpha (q,\omega) &=&
	{\overline{\sum_{M_i}}} \sum_f   \langle \Psi_i | O_\alpha ^\dagger ({\bf q}) | \Psi_f \rangle
		 \langle \Psi_f |  O_\alpha ({\bf q}) | \Psi_i \rangle \nonumber\\
		 &\times&\delta(E_f - E_i - \omega)\ ,\quad\quad \alpha=L,T
		 \label{eq:resp}
\end{eqnarray}
where $O_L({\bf q})=\rho({\bf q})$ is the nuclear electromagnetic charge and $O_T({\bf q})={\bf j}({\bf q})$ is the nuclear electromagnetic current. Here, $| \Psi_i \rangle$ and $| \Psi_f \rangle$  represent, respectively, the initial ground state
and final continuum state with energies $E_i$ and $E_f$, and an average over the initial spin projections $M_i$ of the initial nuclear state with spin $J_i$ (indicated by the overline) is implied. Note that, as $\theta \rightarrow 180^{\circ}$, the double-differential cross-section of Eq.~(\ref{eq:xsec}) is dominated solely by the transverse response function.

The nuclear response functions defined above carry all the information on the nuclear dynamics at play during the scattering event. The electromagnetic charge and current operators are determined by the probe and exhibit dependence upon, {\it e.g.}, the orientation of the nucleons' spins and isospins. Nuclear wave functions, responses, and response densities are calculated within a microscopic model of the nucleus using QMC computational methods~\citep{Carlson:2014vla} to solve the many-body nuclear problem. Within this approach, static and dynamical nuclear properties emerge from the interactions (or correlations) among all the constituent nucleons. For example, nuclear 
responses result from the coupling of external leptonic probes with individual nucleons (described by one-body operators), and with pairs of interacting or correlated nucleons (described by two-body operators).

This scheme  can be appreciated by rewriting the response of Eq.~(\ref{eq:resp}) as
\begin{eqnarray}
 R_\alpha ({\bf q},\omega) &=&
  \int_{-\infty}^\infty  \frac{d t}{2 \pi} \,
  {\rm e}^{ i \left(\omega+E_i\right)   t }\, \nonumber \\
  &\times& \overline{\sum_{M_i}} \,\langle \Psi_i |
 	  O_\alpha^\dagger ({\bf q})\,{\rm e}^{-i H t}\,
 	  O_\alpha ({\bf q}) |\Psi_i \rangle \ ,
\label{eq:realtime}
\end{eqnarray}
where we have replaced the sum over the final states with a real-time propagator. In the equation above, the many-body nuclear Hamiltonian, $H$, consists of single-nucleon (nonrelativistic) kinetic energy terms, and two- and three-nucleon interactions, such that
\begin{equation}
H = \sum_i -\frac{\hbar^2}{2m} \,{\bm \nabla}_i^2+ \sum_{i<j} v_{ij} + \sum_{i<j<k} V_{ijk} \ ,
\end{equation}
where $v_{ij}$ and $V_{ijk}$ are highly sophisticated potentials~\citep{Carlson:2014vla,Bacca:2014tla} which correlate nucleons in pairs and triplets. In the set of calculations used in this work, the Argonne-$v_{18}$ two-nucleon potential~\citep{Wiringa:1994wb} was utilized in combination with the Illinois-7 three-nucleon force~\citep{doi:10.1063/1.2932280}. We indicate this nuclear many-body potential with ``AV18+IL7". The Argonne-$v_{18}$~\citep{Wiringa:1994wb} is a highly sophisticated two-nucleon interaction, reflecting the rich structure of the nucleon-nucleon force, and is written in terms of operatorial structures involving space, momentum, spin and isospin nucleonic coordinates, predominantly arising from one- and two-meson-exchange-like mechanisms. The long-range part of the nucleon-nucleon interaction is due to one-pion-exchange; the intermediate-range component involves operatorial structures arising from multipion-exchange supported by phenomenological radial functions; the short-range part is described in terms of Woods-Saxon functions~\citep{Carlson:1997qn,Carlson:2014vla,Wiringa:1994wb}. The Argonne-$v_{18}$ has 40 parameters that have been adjusted to fit the Nijmegen $pn$ and $pp$ scattering data base~\citep{Stoks:1993tb}, consisting of $\sim 4300$ data in the range of $0–350$ MeV, with a $\chi^2$/datum close to one. While fitting data up to 350 MeV, the Argonne-$v_{18}$ reproduces the nucleon-nucleon phase shifts up to $\sim1$ GeV, an indication that its regime of validity goes beyond the energy range utilized to constrain the adjustable parameters. This is also an indication that relativistic effects are largely embedded in the parameters entering the nucleon-nucleon interaction. The Illinois-7~\citep{doi:10.1063/1.2932280} is the three-body force, supplementing the Argonne-$v_{18}$; its latest formulation involves five parameters constrained (in combination with the Argonne-$v_{18}$) to reproduce $\sim 20$ energy levels of nuclear ground and excited states.

Calculations based on the AV18+IL7 many-body nuclear Hamiltonian successfully explain, both qualitatively and quantitatively, many nuclear electroweak properties~\citep{Bacca:2014tla,Carlson:1997qn,Carlson:2014vla}, including electromagnetic moments and form factors~\citep{Carlson:2014vla,Schiavilla:2018udt,NevoDinur:2018hdo}, low-energy transitions including beta decays~\citep{Pastore:2009is,Girlanda:2010vm,Pastore:2011ip,Pastore:2012rp,Datar:2013pbd,Pastore:2014oda,Pastore:2017uwc,King:2020wmp}, and electron scattering~\citep{Pastore:2019urn}.

The charge, $\rho({\bf q})$, and current, ${\bf j}({\bf q})$, operators are also written as sums of one- and two-nucleon
terms~\citep{Carlson:1997qn,Bacca:2014tla}
\begin{equation}
O_\alpha({\bf q})= \sum_i  O^{(\alpha)}_i({\bf q}) + \sum_{i<j}  O^{(\alpha)}_{ij} ({\bf q})+ \cdots \ .
\end{equation}
Here, we include up to two-body contributions, that is up to operators of the form $O^{(\alpha)}_{ij} ({\bf q})$, where $i$ and $j$ designate that the operator is acting on nucleons $i$ and $j$. The one-body charge and current operators are obtained by taking the nonrelativistic limit of the standard covariant nucleonic currents \citep{Bacca:2014tla,Carlson:1997qn,Carlson:2014vla}, and are written in terms of the nucleonic form factors required to correctly reproduce fall-off at increasing values of three-momentum transfer. In the calculations used in this work, we adopted the dipole parameterization for the proton electric and magnetic, and neutron magnetic form factors, and the Galster form of the neutron electric form factor~\citep{Shen:2012xz}. Other parameterizations or calculations of the nucleon form factors, for example the $z$-expansion~\citep{Meyer:2016oeg}, or calculations from lattice gauge theory \citep{Jang:2019jkn,Rajan:2017lxk,Ishikawa:2018rew,Shintani:2018ozy,Alexandrou:2017ypw,Hasan:2019noy,Alexandrou:2018sjm} can be rather easily implemented within the QMC STA framework.

The two-body currents, ${\bf j}_{ij}({\bf q})$, used in this work have been summarized in \citep{Bacca:2014tla,Carlson:1997qn,Carlson:2014vla} and the references therein. They consist of model-independent and model-dependent terms, the former being constructed by requiring they satisfy the current conservation relation within the Argonne-$v_{18}$. In this sense, they are consistent with the nucleon-nucleon interaction, in that their behaviour at both short and long ranges is consistent with that of the potential, or, equivalently, of two-nucleon correlations. At large internucleon distances, where the nucleon-nucleon interaction is driven by one-pion-exchange, these currents include the standard seagull and pion-in-flight currents. In the seagull mechanism, the external electromagnetic field couples with a nucleon producing a pion which is reabsorbed by a second nucleon, whereas for the pion-in-flight contribution the external field couples to the pion actively being exchanged by two nucleons. The model-independent currents are longitudinal, {\it i.e.}, they are parallel to the direction of the three-momentum transfer ${\bf q}$. The model-dependent two-body currents are orthogonal to the external momentum transferred, and they therefore cannot be constrained using current conservation. The model-dependent dominant term is associated with the excitation of intermediate (virtual) $\Delta$-isobars; in this type of contribution, the external probe excites the nucleon to a $\Delta$ which then decays, emitting a pion which is reabsorbed by another nucleon~\citep{Marcucci:2005zc,Schiavilla:1992sb}. The two-body charge operator, ${\rho}_{ij}({\bf q})$, consists of contributions of one-pion range, which can be regarded as relativistic effects. The specific form of the operators are listed, {\it e.g.}, in \citep{Carlson:1997qn,Marcucci:2005zc}.

Calculations based on the AV18+IL7 two- and three-nucleon correlations in combination with one- and two-nucleon electromagnetic charge and current operators successfully explain available data over a wide range of energy and momentum transfers~\citep{Bacca:2014tla,Carlson:2014vla}. In particular, these calculations highlight the importance of accounting for many-body dynamics---especially two-nucleon dynamics---to achieve agreement with the available experimental data. For example,
corrections from two-body electromagnetic currents enhance the magnetic moments of $^9$C by $\sim40\%$~\citep{Pastore:2012rp}, and give a $\sim 20-40\%$ contribution to both electromagnetic transitions between low-lying nuclear states~\citep{Pastore:2012rp} and electromagnetic transverse response functions~\citep{Pastore:2019urn,Lovato:2017cux}. It is important to emphasize that two-nucleon terms in both the interactions and
currents---collectively indicated by ``two-body physics''---are dominated by one-pion-exchange dynamics.

\subsection{Semifinal States, the Short-Time Approximation, and Response Densities}
Quantum Monte Carlo computational methods~\citep{Carlson:2014vla} have been developed
for the past 30 years to exactly solve the many-body nuclear problem of strongly correlated nucleons. Inclusive response functions, induced by both electrons and $\nu$s, have been calculated in recent years for nuclei up to $^{12}$C~\citep{Carlson:1997qn,Carlson:2001mp,Carlson:1990nh,Lovato:2017cux,Lovato:2015qka,Lovato:2013,Lovato:2014,Lovato:2020kba}.
In particular, one evaluates the Laplace transform of the response~\citep{Carlson:1997qn,Carlson:2014vla}
which results in an imaginary-time response of the type
\begin{equation}
	\widetilde{R}_\alpha ({\bf q},\tau)= \overline{\sum_{M_i}}
   	     \langle \Psi_i |
	  O_\alpha^\dagger ({\bf q}) \,{\rm e}^{-\left( H-E_i\right) \tau}\
	  O_\alpha ({\bf q}) | \Psi_i \rangle\ ,
\label{eq:imaginarytime}
\end{equation}
where Green's function Monte Carlo (GFMC) methods can then be used to calculate the relevant matrix elements between  ground-state wave functions~\citep{Carlson:2014vla}.
Since the nuclear response in the QE region is fairly smooth as a function of $\omega$,
maximum entropy techniques are successful in inverting the Laplace transform to obtain the
response function~\citep{Lovato:2015qka}. Within this scheme, one can fully account for the correlations
in the initial state and the interaction effects induced by the imaginary time  propagator into the final state,
along with quantum interference effects. Interference between one- and two-body currents plays a crucial role
in explaining the experimentally observed enhancement in the electromagnetic transverse responses function~\cite{Carlson:2001mp} and should not be
neglected in calculations of nuclear responses.

While being extremely successful in explaining available scattering data, the GFMC approach is computationally costly, which is why it has been applied only to nuclei up to $A=12$. To meet the demands of the next generation neutrino oscillation experiments that will be utilizing $^{40}$Ar as active material in the detectors, one has to resort to approximated computational schemes to calculate the associated nuclear responses. The STA~\cite{Pastore:2019urn} has been developed to address this issue without losing the resolution acquired by the exact GFMC calculations, that is, without losing the important contributions from two-body correlations and electroweak currents. The STA is based on the factorization of the real time response given in Eq.~(\ref{eq:realtime}) at short-times (high-energies). In particular, only one- and two-body terms in the Hamiltonian entering the real time propagator are kept. The STA then fully retains two-body physics from both the Argonne $v_{18}$ and the associated electromagnetic one- and two-body currents, and resultant interference terms. The initial state wave functions are fully correlated, as in the GFMC case. When used to calculate response functions, the STA produces results that are in very good agreement with the GFMC calculations at high energy transfers, $\omega$, and moderate to high values of momentum transfer, ${\bf q}$. The low energy behaviour induced by low-lying nuclear excitations and by collective excitations are not captured by the STA.

In this foundational work, we implement only these electromagnetic response functions into the generator. However, the STA, due to the factorization scheme, provides us with additional important information on the leptonic \textit{and} hadronic ``\textit{semi}final" states--in particular, for two-nucleon semifinal states struck by the external probe via one- and two-body electroweak currents before transport through the nuclear medium. This information is cast in nuclear response \textit{densities}, $\mathcal{D}(e, E_{\rm c.m.})$, which are expressed in terms of the relative ($e$) and center of mass ($E_{\rm c.m.}$) energies of the struck nucleon pair (or equivalently in terms of the relative and center on mass momenta of the pair). Upon integration of the response densities,  one recovers the response functions via
\begin{eqnarray}
 \label{eq:density.formula.e}
R^{\rm STA}_{\alpha}({\bf q},\omega)& =& \int_0^\infty \, de  \int_0^\infty  dE_{\rm c.m.} \, \,
 \mathcal{D}_{\alpha}(e, E_{\rm c.m.}) \nonumber \\
&\times& \delta \left(\omega+E_i-e - E_{\rm c.m.} \right)   \ ,
\end{eqnarray}
where for simplicity we have ignored the Jacobian.

\begin{figure}[h]
    \centering
    \includegraphics[height=2.2in]{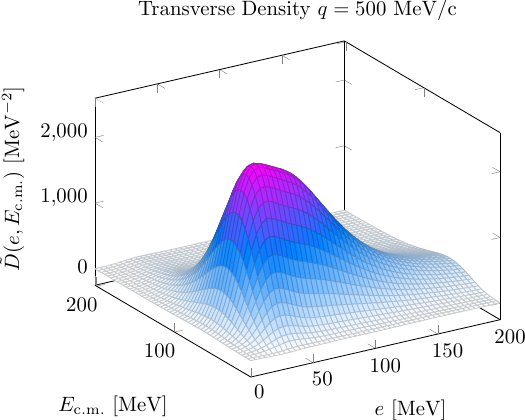}
    \caption{The \isotope[4]{He} transverse response density is shown for ${\bf q}=500\,$MeV/c. The surface plot shows the response density as functions of relative energy $e$ and center-of-mass energy $E_{\rm c.m.}$ of pairs of nucleons being actively scattered upon by the incoming electron, leading to microscopic knowledge of \textit{semi}final states before intranuclear transport and final state interactions.}
    \label{fig:3dplot}
\end{figure}

The transverse response density induced by  electrons scattering from $^4$He is displayed in Fig.~\ref{fig:3dplot}. The implementation of this semifinal hadronic state information at the interaction vertex within {\fontfamily{qcr}\selectfont GENIE} will be the subject of a further work currently in preparation.

\section{{\fontfamily{qcr}\selectfont GENIE} Implementation}

Steps have been taken within the {\fontfamily{qcr}\selectfont GENIE} collaboration to create a new suite of software tools to allow for external contributors to implement their inclusive cross section calculations in a universal way using tabulated nuclear responses. Interpolation of these responses allows for the calculation of double differential cross sections at various kinematics, permitting validation against experimental data. Given the usually large $|{\bf q}|$-spacing between known responses, the sensitivity of these cross sections to the interpolation method can be nontrivial, occasionally leading to discontinuous behavior; secondarily, given a particular calculation's legitimacy within certain energy regimes (and the limitations of computational time and tabulated data sets), one may not be able to continuously interpolate cross sections to \textit{all} conceivable kinematic regimes. However, creating a fine grid over a legitimate QE kinematic regime permits one to reduce each of these unsavory effects. Here, we discuss some of these solutions in more detail.

\subsection{Cross section calculation}

To facilitate implementation of new lepton-nucleus cross section calculations, the \texttt{GENIE} collaboration has developed an interface for pre-computed nuclear responses to be used in event generation. The technique relies on the observation that the inclusive differential cross section can be written very generally in the form
\begin{equation}
\label{eq:xsec_tensors}
    \frac{d^2\,\sigma}{d\,\omega d\,\Omega}= \frac{\mathcal{C}}{\pi^2}\, \frac{|\mathbf{k}^\prime|}{|\mathbf{k}|}\,L_{\mu\nu}\,W^{\mu\nu}\, ,
\end{equation}
where $\mathbf{k}$ ($\mathbf{k}^\prime$) is the initial (final) three-momentum of the lepton, $L_{\mu\nu}$ ($W^{\mu\nu}$) is the
leptonic (hadronic) tensor, and
\begin{equation}
\label{eq:coupling_factor}
\mathcal{C} \equiv
\begin{cases}
  \frac{1}{2} \, G_F^2 \, |V_\mathrm{ud}|^2 & \text{CC processes} \\
  \frac{1}{2} \, G_F^2 & \text{NC processes} \\
  \frac{\alpha^2}{Q^4} & \text{EM processes} \\
\end{cases}
\end{equation}
is a factor that contains the coupling constants appropriate for the scattering process of interest. For Standard Model processes, the leptonic tensor is well-known and given by a trace over Dirac matrices. The elements of the hadronic tensor may be computed in terms of nuclear response functions. Exploiting the Lorentz invariance of the tensor contraction $L_{\mu\nu}\,W^{\mu\nu}$, \texttt{GENIE} evaluates these in a frame in which the three-momentum transfer $\mathbf{q}$ points along the $+z$ direction. For electromagnetic scattering in such a frame, contributions from only two elements of $W^{\mu\nu}$ are nonvanishing:
\begin{align}
W^{tt} &= R_L,\\
W^{xx} &= R_T,
\end{align}
where the nuclear responses $R_L$ and $R_T$ are defined as in Section~\ref{sec:QEInclusiveXSec}.

Pre-computed tables of nuclear responses, evaluated on a two-dimensional grid in $(\omega,|\mathbf{q}|)$ space, may be provided to \texttt{GENIE} as a set of text files organized by target nucleus and interaction mode (e.g., a table may include only the one-body contribution). A simple nearest-neighbors bilinear interpolation scheme is used to evaluate the hadronic tensor elements $W^{\mu\nu}$ between the grid points. The numerical results obtained in this way are used to evaluate inclusive double differential cross sections using the standard form of the leptonic tensor $L_{\mu\nu}$. Further implementation details are available in ref.~\citep{HadTensorNote}.

The \texttt{GENIE} strategy described above for inclusive cross section calculations originated in work to implement the Valencia model \cite{ValenciaMEC1,ValenciaMEC2} for CCMEC interactions \citep{GENIEValenciaMEC}. The treatment used therein was subsequently generalized and improved to allow for its application to other scattering processes (e.g., EM interactions). In addition to the model presented here, the same code framework was also recently used to add the SuSAv2 calculation \citep{Amaro:2019zos,Dolan:2019bxf} of QE and MEC cross sections to \texttt{GENIE} for both neutrinos \citep{SuSAv2Validation} and electrons \citep{Papadopolou2020}.

\subsection{Scaling and Interpolation Techniques}

Given the computational difficulty in directly evaluating the STA nuclear responses on a finely-spaced ($\sim1\,$MeV) grid in $(\omega,|{\bf q}|)$ space, one must employ \textit{one among many} possible and legitimate forms of interpolation on the available sparse $\{R,\omega,|{\bf q}|\}$ surface \citep{Pastore:2019urn}. For practical calculations in an event generator, the interpolation method must be fast and efficient while avoiding storage of very large tables in memory. The ability to handle input files for which the $\omega$ and $|\mathbf{q}|$ grid points are not regularly spaced is also highly desirable.

All of the above is accomplished within the {\fontfamily{qcr}\selectfont GENIE} MC$\nu$EG using a recently-developed ``hadron tensor'' interface \citep{HadTensorNote}, which computes cross sections using bilinear interpolation to obtain nuclear response values between grid points. For the \isotope[4]{He} EM responses used in this study, the input tables use a spacing of 2~MeV between $\omega$ grid points and 1~MeV between $|{\bf q}|$ grid points. The kinematic limits of the grid are 1 MeV $\leq |{\bf q}| \leq 2000$ MeV and 2~MeV $\leq \omega \leq$ 1800~MeV.

\begin{figure}[h]
    \centering
    \includegraphics[width=1.0\columnwidth]{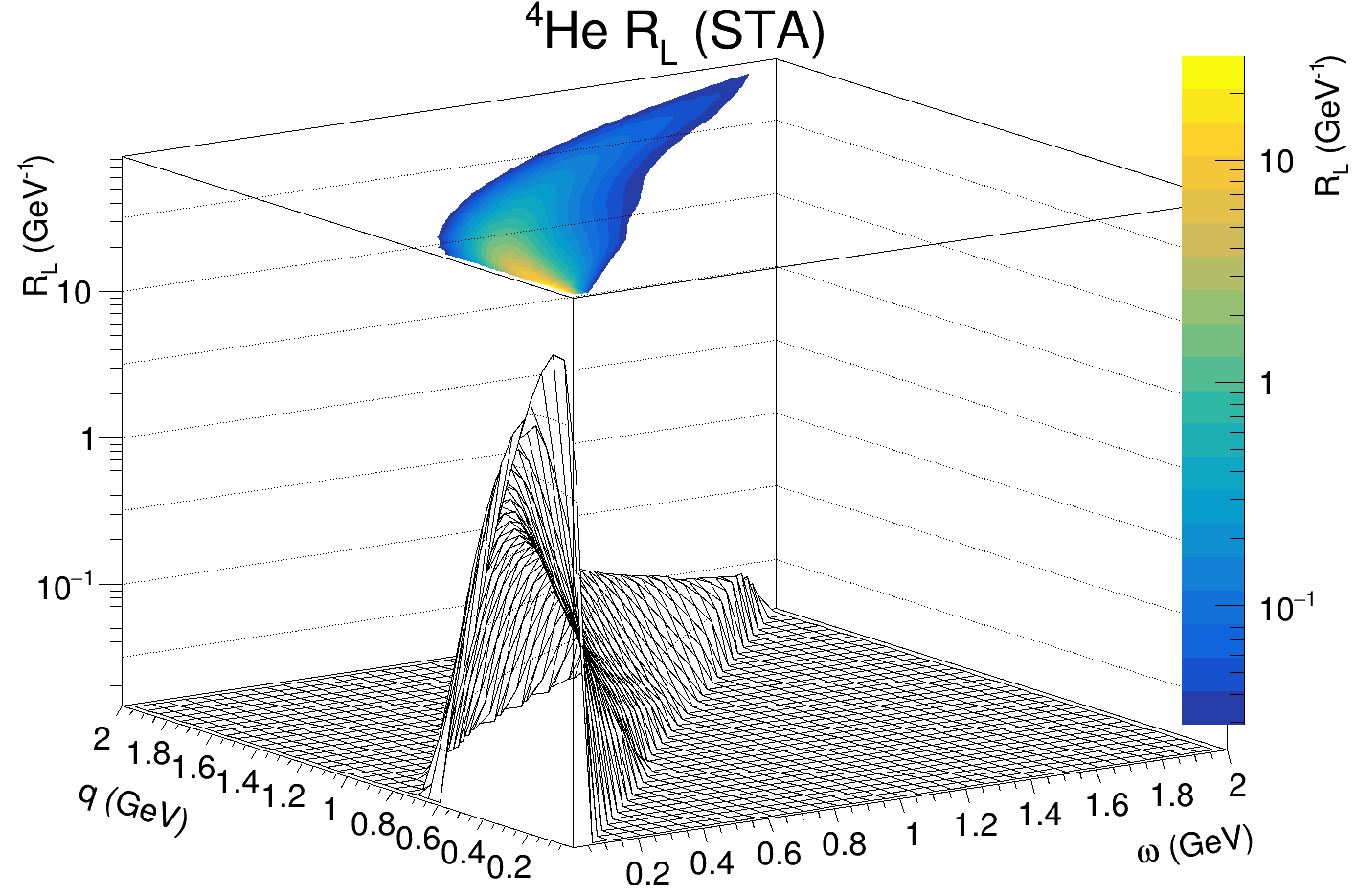}
     \includegraphics[width=1.0\columnwidth]{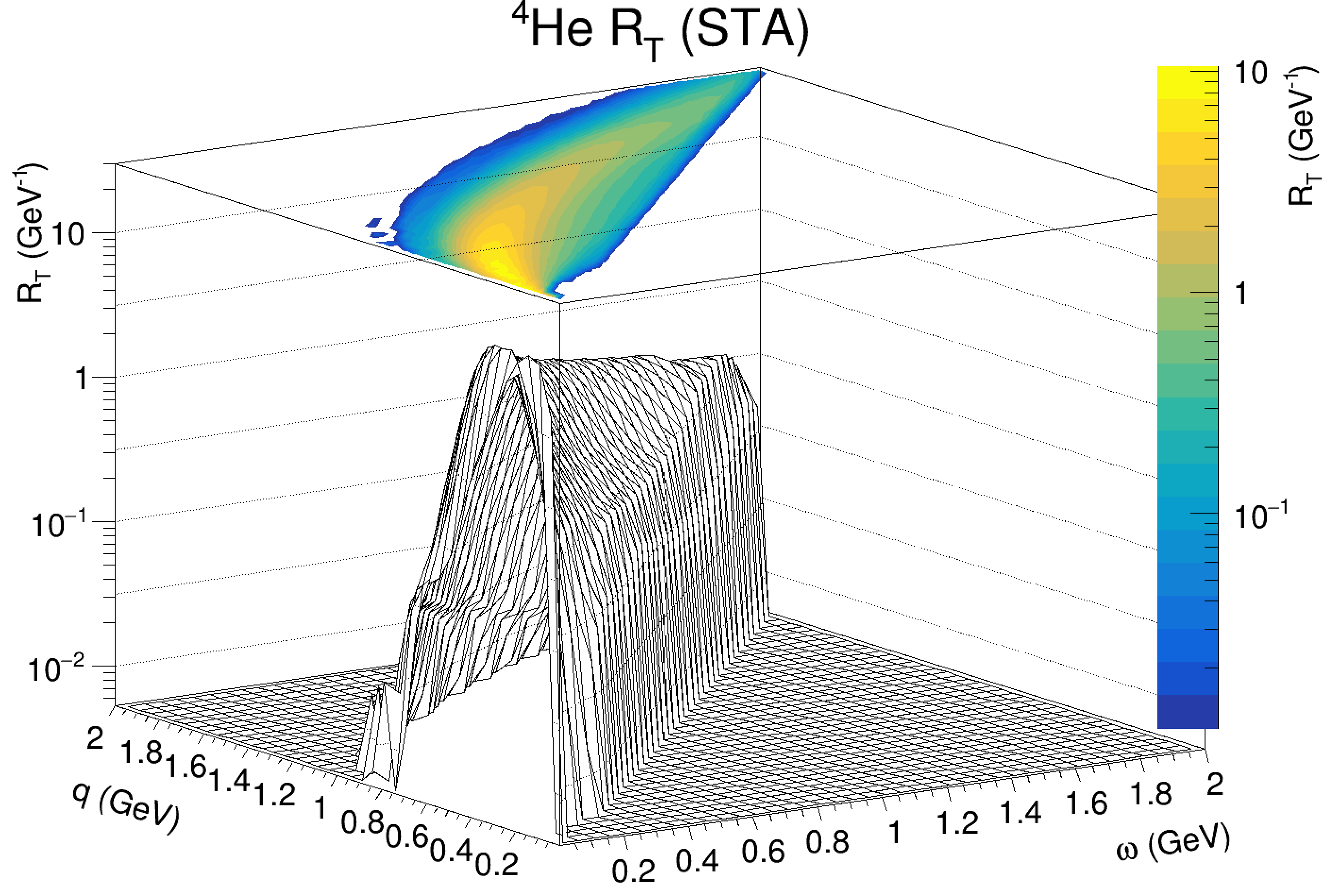}
    \caption{The interpolated nonrelativistic nuclear response surfaces $\{R_{\alpha},q,\omega\}$ are shown with sub-MeV grid-spacing. The underlying $\sim 1\,$MeV-spaced $\{|{\bf q}|,\omega\}$ grid forms the fundamental objects cast in tabulated form which {\fontfamily{qcr}\selectfont GENIE} then dynamically bilinearly interpolates upon to form all subsequent double differential cross sections for QE EM scattering. Lines along the surfaces serve as visual aides only.}
    \label{fig:ResponseInterpolation}
\end{figure}

Currently, we employ only one of the many potential techniques one could use to achieve such a high granularity on the $\{|{\bf q}|,\omega\}$ grid \textit{with} good accuracy.  We choose to use an approximately $|{\bf q}|$-invariant object, a nonrelativistic \textit{scaling function}, to make thousands of \textit{new} nonrelativistic total nuclear response functions at many \textit{new} momentum transfers. These objects are created in a one dimensional way; other future methods will be able to utilize the full multidimensional nature of the response \textit{densities} \citep{Pastore:2019urn}, and will be discussed further in the Conclusions and the Appendix. These scaling functions can be calculated from one among several existing nonrelativistic nuclear response functions \citep{Pastore:2019urn}, in-turn creating a single \textit{average} nonrelativistic scaling function $\overline{f^{nr}_{\alpha}}[\psi^{nr}(|{\bf q}|,\omega)$ \citep{Donnelly_1999,Rocco:2017hmh,Rocco:2018tes,Lovato:2020kba} built up from any \textit{set} of scaling functions, $f_{\alpha,i}^{nr}$, as follows:
\begin{center}
    \begin{equation}
        f_{\alpha,i}^{nr}[\psi^{nr}(|{\bf q}_{i}|\in\widetilde{Q},\omega)] = k_F \cdot \frac{R_{\alpha}^{nr}(|{\bf q}_{i}|\in\widetilde{Q}),\omega)}{G_{\alpha}^{nr}(|{\bf q}_{i}|\in\widetilde{Q})},
    \end{equation}
    \begin{equation}
        \therefore\overline{f^{nr}_{\alpha}}[\psi^{nr}(|{\bf q}|,\omega)] = \frac{1}{N} \sum^N_{i=1} f_{\alpha,i}[\psi^{nr}(|{\bf q}_{i}|\in\widetilde{Q},\omega)],
    \end{equation}
    \begin{equation}
        \longrightarrow R_{\alpha}^{nr}(|{\bf q}|,\omega)=\frac{1}{k_F}\cdot G_{\alpha}(|{\bf q}|) \cdot \overline{f^{nr}}[\psi^{nr}(|{\bf q}|,\omega)],
    \end{equation}
\end{center}
where $\psi^{nr}\equiv\psi^{nr}(|{\bf q}|,\omega)$ is a \textit{nonrelativistic} scaling variable \citep{Donnelly_1999,Lovato:2020kba}, $R_{\alpha}^{nr}(|{\bf q}_{i}|\in\widetilde{Q},\omega)$ is a known nonrelativistic nuclear response function for a particular component $\alpha$ from a known computed set \citep{Pastore:2019urn} of momentum transfers $|{\bf q}|\in\widetilde{Q}=\{400,450,\ldots,750,800,1000\}\,$MeV and where $\widetilde{Q}$ is of size $N$, $G_{\alpha}^{nr}(|{\bf q}|)$ can be any component-specific functional combination of single nucleon electric and magnetic form factors, and $k_F$ is the nominal Fermi momentum of the system. The final $\{R_{\alpha},|{\bf q}|,\omega\}$ surfaces resultant from this averaged scaling shown in Figs. \ref{fig:ResponseInterpolation} serve as the basis objects for all double differential cross section calculations in {\fontfamily{qcr}\selectfont GENIE}, and are displayed after subsequent sub-MeV bilinear interpolation on the tabulated grid.

\onecolumngrid

\twocolumngrid

Note that Ref.~\citep{Lovato:2020kba} shares a common theoretical basis with inputs used in this work \citep{Pastore:2019urn}, and also shows that good scaling behavior persists \textit{even with the inclusion of two-body dynamics}. When comparing to the originally computed longitudinal nuclear response functions, this method \textit{partially} removes some endemic contamination of the \textit{elastic} scattering component, which would otherwise lead to over-estimations of longitudinal response function at low momentum transfers ($q \lesssim 300$ MeV). However, as we will show, the use of an averaged scaling function takes away too much strength from the double differential cross section in some kinematics (particularly those of high outgoing angles), the origin of which is the averaging process itself $q$ (see Appendices, \ref{sec:Appendix}, Fig. \ref{fig:1BTotNewvsOldResponseFunctions} to see these differences). In particular,  the averaging technique reduces the strength of {\it both} response functions at low ${ q}$---as shown in Fig.~\ref{fig:1BTotNewvsOldResponseFunctions}, indicating that this technique needs to be improved to achieve a good agreement with the data in the aforementioned kinematic regimes.  For a  more detailed overview on this average scaling method, and how it attempts to recreate known responses, see the Appendices in Sec. \ref{sec:Appendix} and the references therein. In closing, we point out that the elastic peak that is currently contaminating the longitudinal responses at low $q$ can be removed directly within the QMC-STA calculations. Work along these lines is underway.

\section{Validation}

\subsection{Inclusive Electromagnetic Responses and Double Differential Leptonic Cross Sections}

\begin{figure}[h]
    \centering
    \includegraphics[width=1.0\columnwidth]{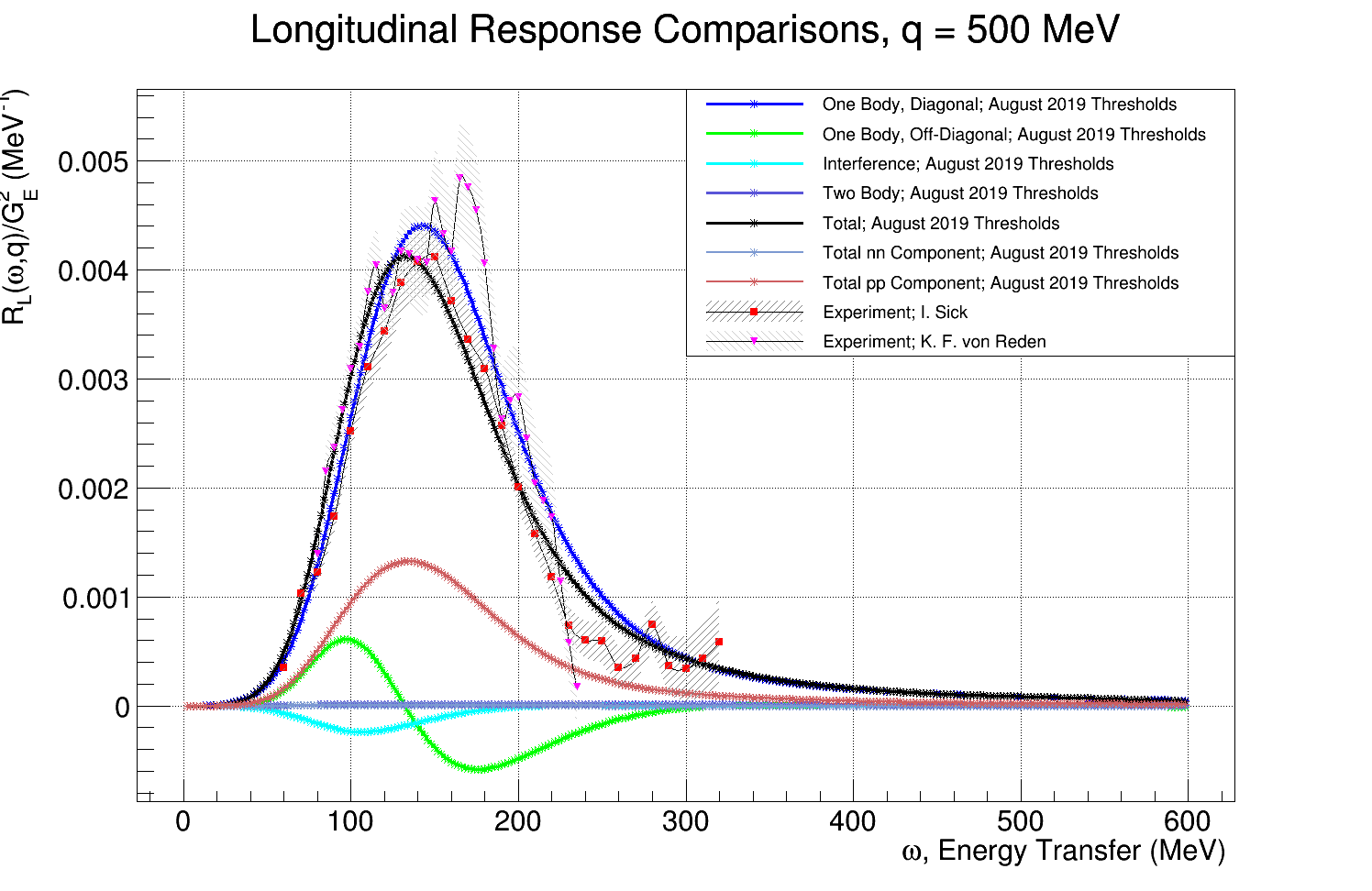}
    \includegraphics[width=1.0\columnwidth]{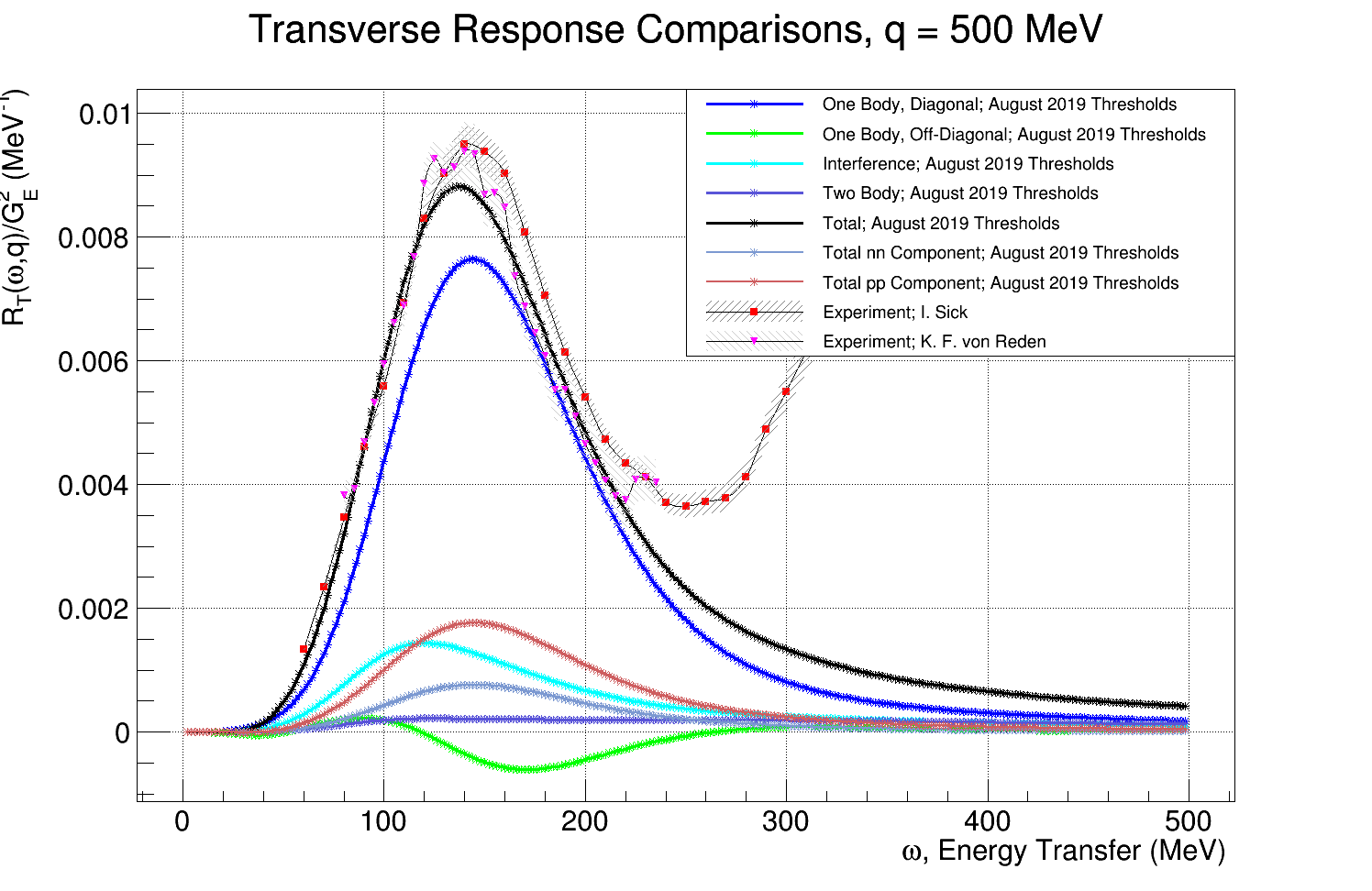}
    \caption{Nuclear responses comparisons between QMC STA theory outputs \citep{Pastore:2019urn} and available empirical data \citep{vonReden:1990ah,Carlson:1997qn}. Many response components are shown, including but not limited to \textit{interference} and one-body \textit{off-diagonal} terms, whose destructive qualities within particularly the longitudinal response limit the strength of the pure one-body contribution. Thresholds refer to a \textit{small}, free shift-parameter which has been simplistically tuned in post-processing to better fit available response data.}
    \label{fig:nuclearresonses}
\end{figure}

Transverse and longitudinal nuclear response functions \citep{Pastore:2019urn} have been validated against available EM nuclear response data sets from Sick \textit{et al.} \citep{Carlson:1997qn} and von Reden \textit{et al.} \citep{vonReden:1990ah} where possible, as shown in Figs. \ref{fig:nuclearresonses} with excellent agreement without direct pion production.

\onecolumngrid

\begin{figure}[h!]
    \centering
    \includegraphics[width=0.65\columnwidth]{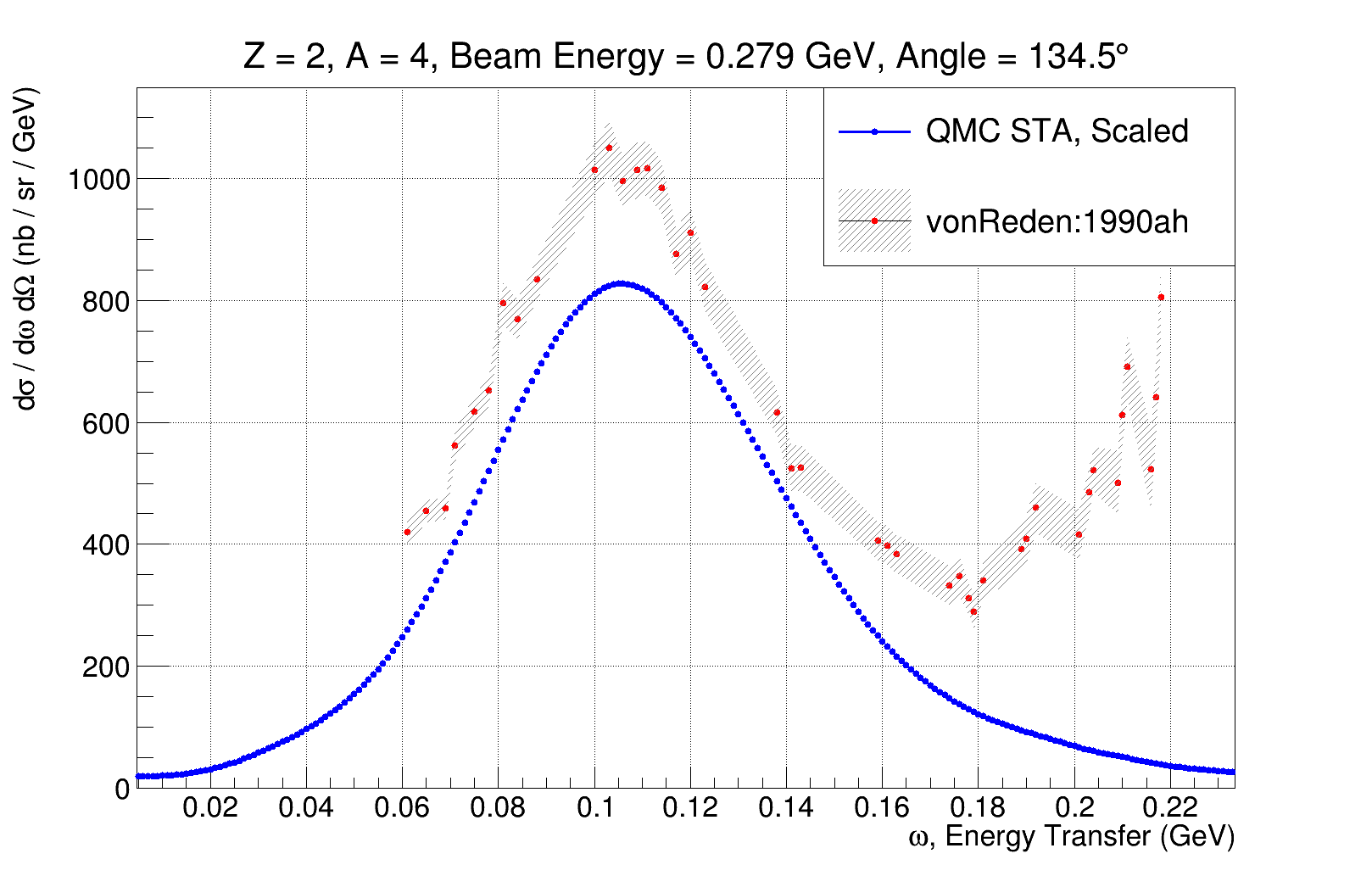}
    \includegraphics[width=0.65\columnwidth]{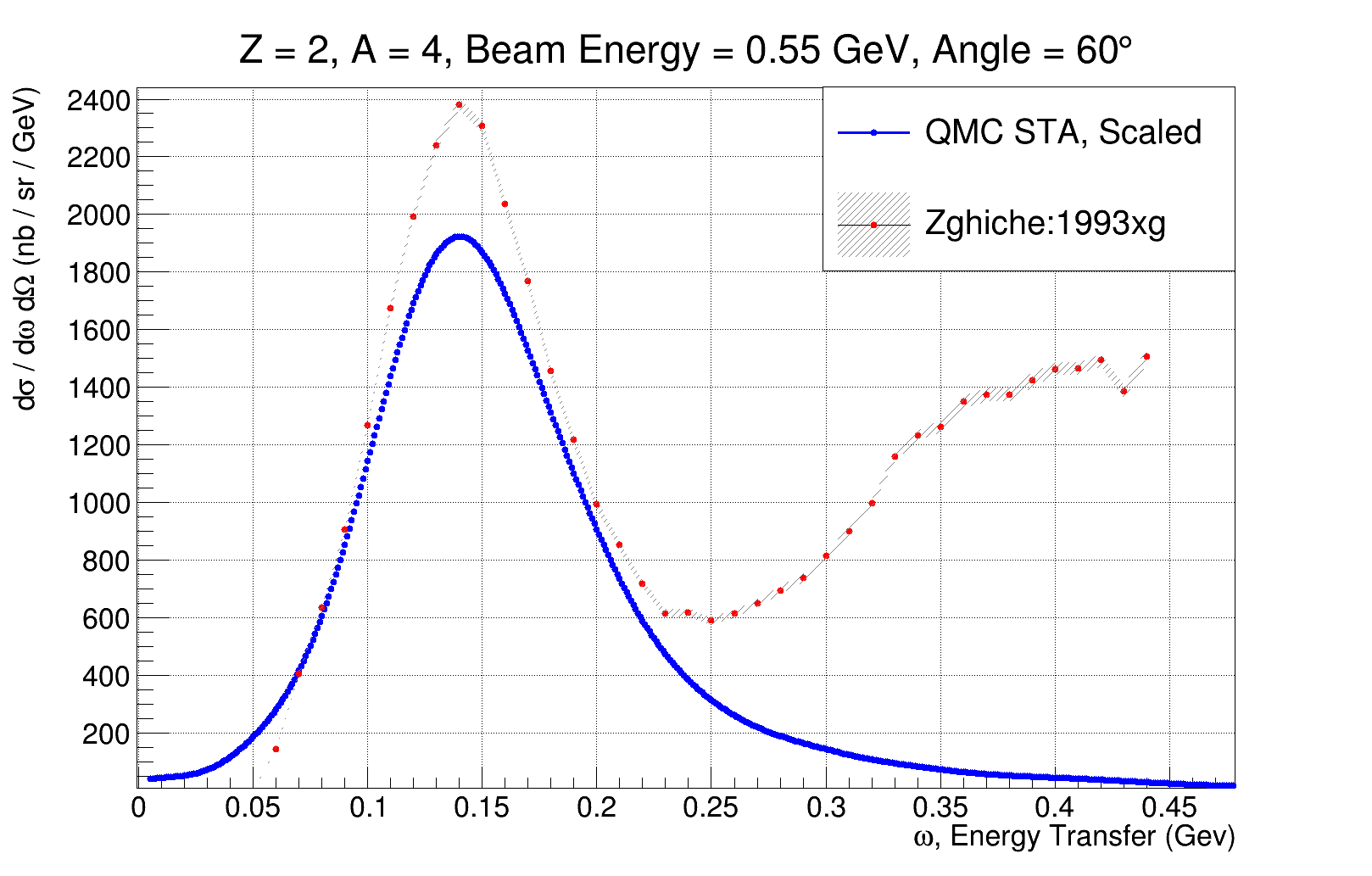}
    \includegraphics[width=0.65\columnwidth]{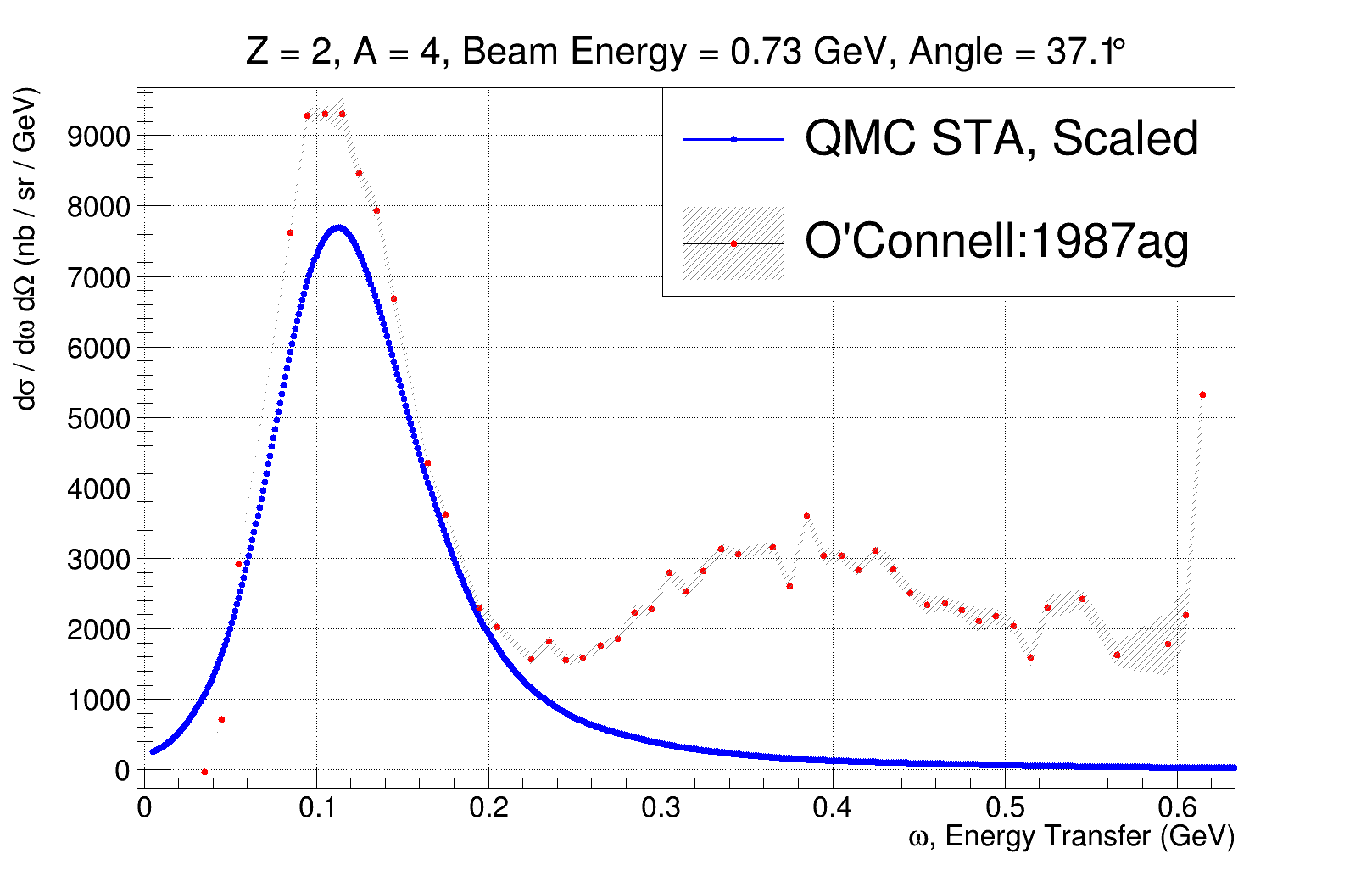}
    \caption{A series of \isotope[4][2]{He}~double differential leptonic cross sections are shown for various beam energies and angles, derived from the scaled responses coming from the average scaling function. Behavior is good overall, with all curves properly and consistently undershooting the QE-peak due to lack of resonant production. This is especially true for beam energies $<2\,$GeV and more forward angles, though even highly transverse cross sections appear quite consistent with data. However, one can see that strength is missing from the top-most plot at low energy and high angle, due to the current averaging scheme of scaled nuclear responses.}
    \label{fig:QEXSectionsPart1}
\end{figure}

\twocolumngrid
A tool utilizing {\fontfamily{qcr}\selectfont GENIE}'s hadron tensor framework completes bilinear interpolation of scaled nuclear responses across $|{\bf q}|$ and $\omega$, allowing for calculation of double differential cross sections from scaled nuclear response inputs, and so one may compare to available world QE double differential cross section data sets \citep{Benhar:2006wy,EMQEOnlineDataSet} for
\isotope[4][2]{He}.  A small selection of these can be seen in Figs. \ref{fig:QEXSectionsPart1}. This simple technique shows good comparative power to data \textit{despite} the use of the averaging interpolation techniques and the lack of explicit removal of the elastic peak, relativistic corrections, or on-shell $\pi$-production via $\Delta$ resonances, broadly matching the QE position and width up to around $2\,$GeV of electron beam energy (the highest scaled response $|{\bf q}|$-value utilized is $2\,$GeV/c). The full statistical consistency of these model curves with data across all available angles and energies will be pursued in future work; comparison between various model predictions may also be pursued. Thus, as a purely QE theory \citep{Pastore:2019urn}, one observes MC-generated double differential cross sections beneath experimentally determined ones; this is in part thanks to scaling's effective removal of the elastic peak, but also due to the averaging scheme, which lowers the strength of the responses slightly too much at certain kinematics. Cross sections do remain consistent with experimental ones at moderate to high momentum transfers and moderate to high scattering angles, where the transverse response of the nucleus containing two-body dynamics plays a disproportionate role. Once coherent modules are complete for \textit{simultaneous} event generation of leptonic and hadronic variables for the semifinal state at the interaction vertex, thus utilizing the STA response \textit{densities}, it is planned that the scaled response function averaging scheme will be supersceded by another nonlinear multidimensional interpolation technique \citep{Baak:2014fta}; comparisons with current techniques will follow in a future work.

\subsection{Double Differential Leptonic Cross Sections and Approximate Two-Body Final State Predictions}

\begin{figure}[h!]
    \centering
    \includegraphics[width=1.0\columnwidth]{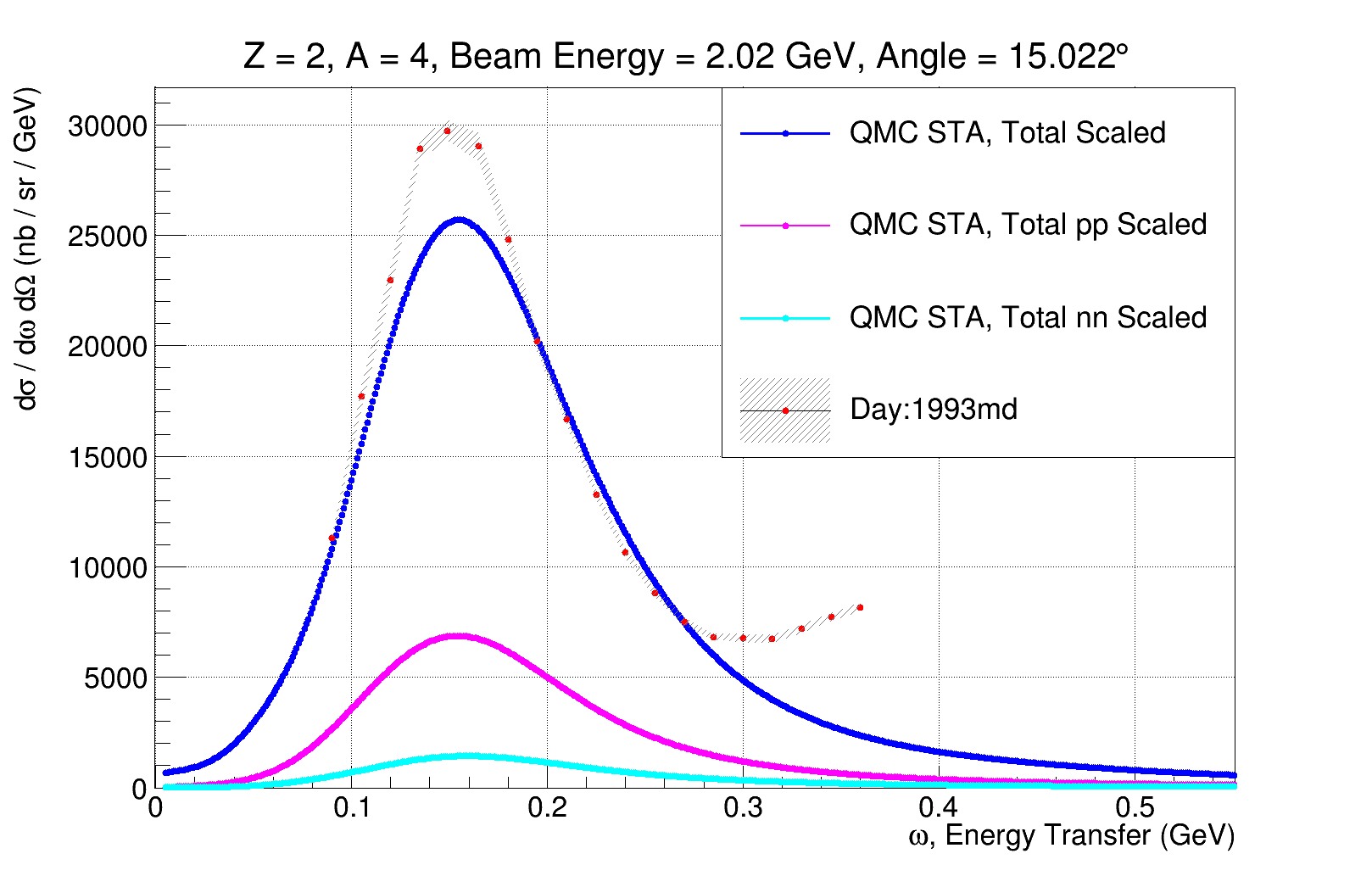}
    \includegraphics[width=1.0\columnwidth]{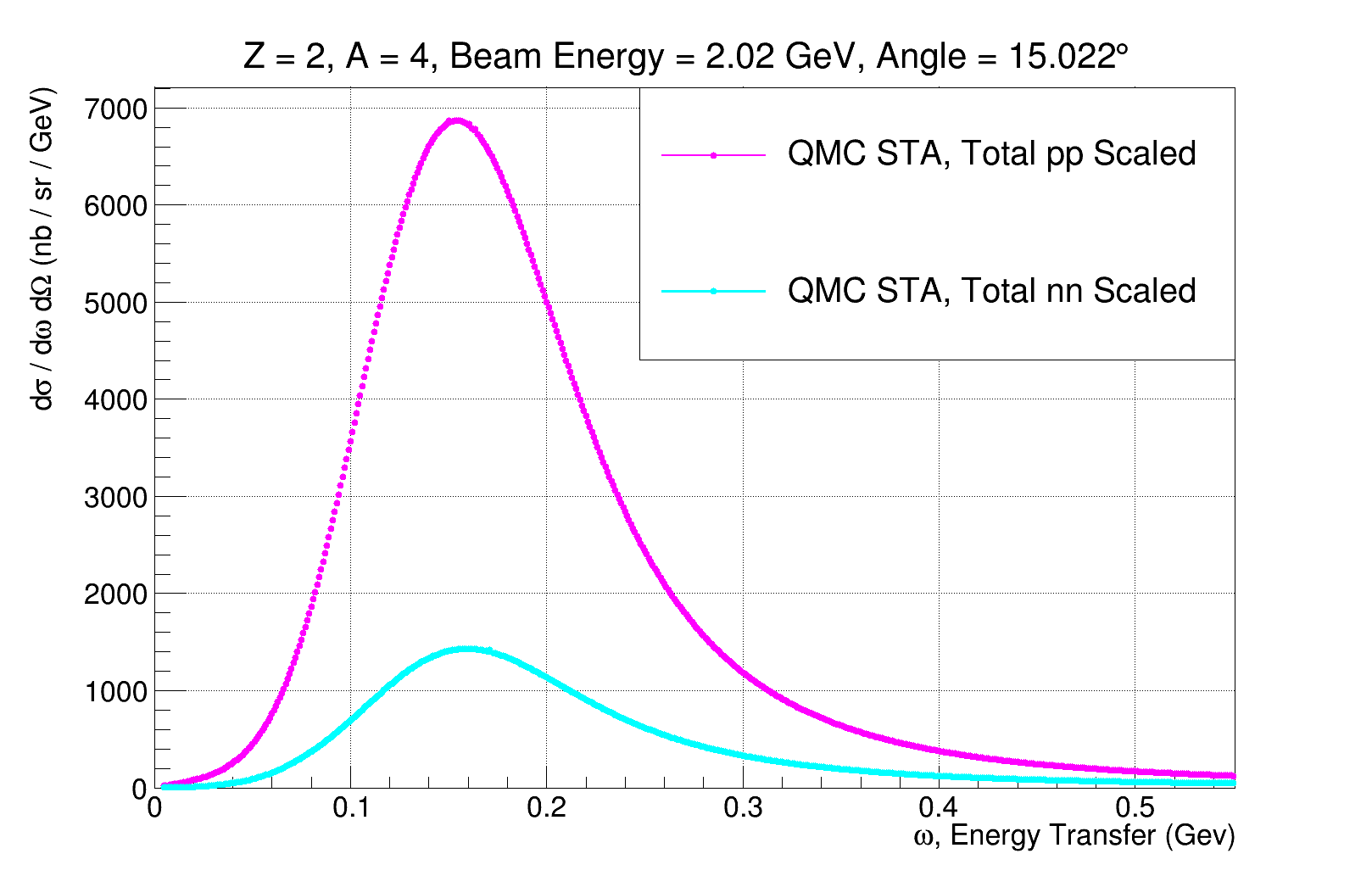}
    \caption{A prediction of total inclusive double differential electron scattering cross sections. The  $pp$ and $nn$ channels are also shown.
    }
    \label{fig:QETwoBodyXSec}
\end{figure}

Using the QMC STA formalism, one may also consider total QE EM (reaction) double differential leptonic cross sections of individual components of the nuclear structure, giving one access to the   one- and two-body contributions. In Figs. \ref{fig:QETwoBodyXSec}, we see the theoretical total QE EM double differential cross section (dark blue) matches the shape and peak position of available data (red) quite well, while again properly underpredicting the total due to lack of $\pi$-production. Individual shapes of the cross sections for scattering from $pp$ (pink) and $nn$ (light blue) pairs can also be seen, including in a zoomed-in view (lower).

\onecolumngrid

\twocolumngrid

The pink and light blue curves shown are derived from an identical scaling method utilizing known $|{\bf q}|=\{ 500,600,700 \}$MeV/c particle identity-specific nuclear response functions.  It should be stressed here that the (lower) plots in Figs. \ref{fig:QETwoBodyXSec} are speaking to the final state lepton \textit{only}; however, such a final state lepton indeed must be \textit{approximately} commensurate with the appearance of $pp$ and $nn$ final states. This is approximate due to the nature of intranuclear FSIs, where multiple scattering can (generally) lead to reductions in the struck nucleons' kinetic energy to potentially below the Fermi energy; the resultant final state topology could then become $eNN\rightarrow eN$. Similarly, again \textit{due} to FSIs, one may potentially have a true QE interaction, but multiple scattering may be such that two nucleons enter the final state, {\it i.e.}, $eN\rightarrow eNN$. The interference of these effects will be studied in greater detail in future work, where marriage between leptonic and hadronic components of the QMC STA will be mediated by correlated use of both QMC STA response \textit{densities} and {\fontfamily{qcr}\selectfont GENIE} FSI models; once complete, two-nucleon final state data will be considered to validate (and \textit{potentially} tune) the full generator module.

\subsection{Generating {\fontfamily{qcr}\selectfont GENIE} Leptonic Events}
Sampling of the lepton kinematic variables is handled by \texttt{GENIE} in the same way as for the SuSAv2 implementation \citep{SuSAv2Validation}. An accept/reject approach is used to select a value of final lepton kinetic energy $T_\ell$ and its scattering cosine $\cos\theta_\ell$ in the lab frame from the probability distribution
\begin{equation}
\label{eq:kinematic_sampling}
P(T_\ell,\cos\theta_\ell) = \frac{1}{\sigma}
\, \frac{d^2\sigma}{dT_\ell \, d\cos\theta_\ell}
\end{equation}
where $\sigma$ is the total cross section. The maximum value of the differential cross section in Eq.~(\ref{eq:kinematic_sampling}), which is needed for rejection sampling, is found via a brute-force scan over the two-dimensional phase space. After $T_\ell$ and $\cos\theta_\ell$ have been selected, a value for the azimuthal scattering angle $\phi_\ell$ is chosen uniformly on the interval $[0, 2\pi)$.

\begin{figure}[h]
    \centering
    \includegraphics[width=1.0\columnwidth]{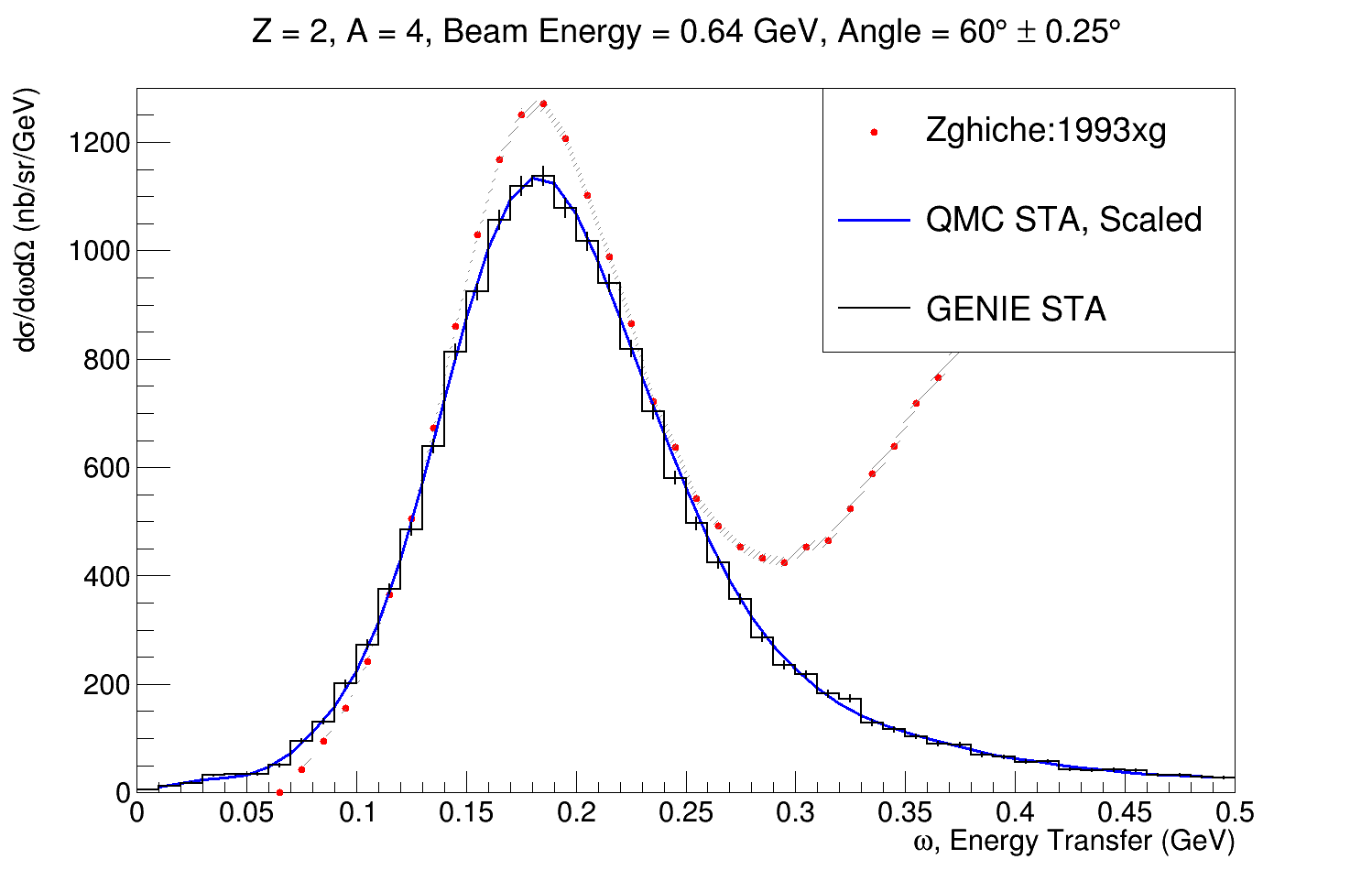}
    \includegraphics[width=1.0\columnwidth]{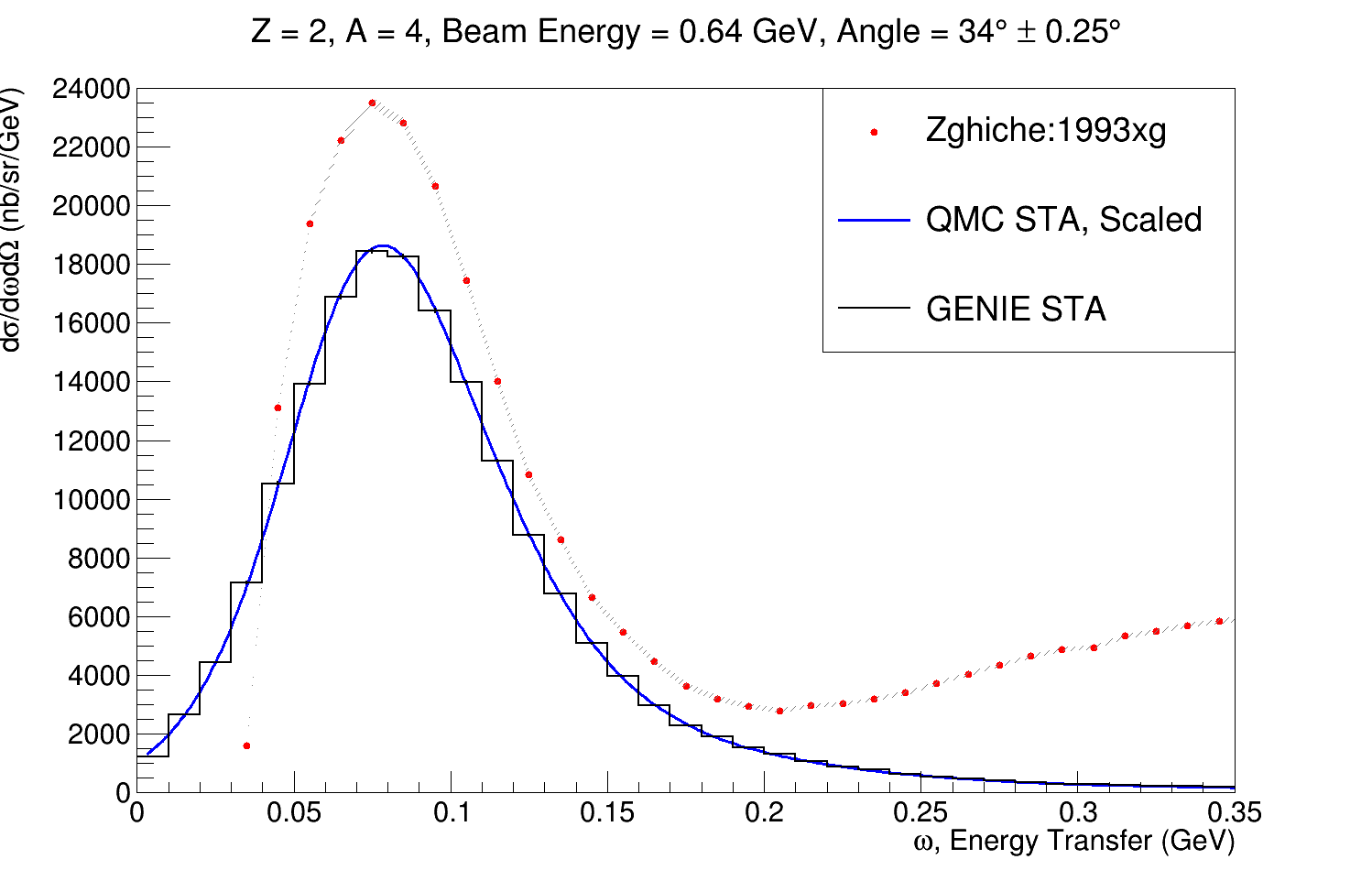}
    \caption{Two kinematics are shown for double differential cross sections showing data, scaled theoretical curves, and {\fontfamily{qcr}\selectfont GENIE} generator outputs. Great consistency in all three is observed throughout the QE regime.}
    \label{fig:QMCSTAGENIE}
\end{figure}

In Fig.~\ref{fig:QMCSTAGENIE} we show representative results from our efforts to validate \texttt{GENIE} simulations of inclusive electron-\isotope[4]{He} scattering using the STA nuclear response functions \citep{Pastore:2019urn} described in this work. In each plot, the measured cross section \citep{Zghiche:1993xg,EMQEOnlineDataSet} at fixed scattering angle is drawn using red points, while the QMC STA calculation is shown by the blue curve. Cross sections computed using \texttt{GENIE} events (with the angular acceptance indicated in the plot title) are drawn as black histograms. Excellent agreement is seen between the generated events and the underlying STA calculation. Some expected strength can be seen to be missing from the lower plot, again due to the averaging scheme.

\onecolumngrid

\twocolumngrid

\section{Conclusions and Outlook}

We have shown the formulation of a new quasielastic electron scattering module for final state leptonic variables conceived within the {\fontfamily{qcr}\selectfont GENIE} Monte Carlo event generator using scaled and tabulated nuclear response function inputs from quantum Monte Carlo Short-Time Approximation calculations. Importantly, the model implemented within our event generator module retains one-body, two-body, \textit{and interference} physics in a fully quantistic manner within the quasielastic scattering regime, a unique and powerful addition to better understand experimental measurements. Despite the marked computational intensity to simulate the many-body problem, and the current contamination of the elastic peak in the calculated  longitudinal response function at low momentum transfer ($q\lesssim 300$ MeV), and thanks to the (approximated)  average scaling  analysis and proceeding interpolations, prodigious world data comparisons to {\fontfamily{qcr}\selectfont GENIE}-derived outputs over a large range of quasielastic momentum transfers show good agreement for both longitudinal and transverse angles, particularly in that no predicted double differential cross sections overshoot experimental data which contain resonant production. Going forward, we will be working toward the direct removal of the elastic peak from the Short-Time Approximation's longitudinal nuclear response functions and densities at low momentum transfer.

We have hinted throughout this article on a more all-encompassing path beyond this work, where instead Short-Time Approximation nuclear response \textit{densities} and their descriptions of intranuclear semifinal states will be eventually married to {\fontfamily{qcr}\selectfont GENIE} final state interaction models for intranuclear transport to assess simultaneous correlations between the outgoing lepton and one-or-two nucleons. The numerical interpolation \citep{Baak:2014fta} between and integration of these densities within {\fontfamily{qcr}\selectfont GENIE} itself, and comparison to \textit{known} Short-Time Approximation outputs, will allow for robust validation amidst ongoing event generation, independent scaling behavior, $Z$, or $A$. If this powerful method shows consistency between data and the resulting generated cross sections, it will be able to supersede the current average scaling function analyses and interpolation schemes shown here, simultaneously generating correlated semifinal state behavior for \textit{both} the lepton and hadrons moving out from the interaction vertex. None-the-less, optimization of scaling behavior for even stronger consistency with data will be pursued for the outgoing lepton, possibly by the use of weighted averaging and $\chi$-square comparisons against data. Also, other nonlinear nearest-neighbor interpolation schemes can be pursued between scaled responses \textit{and} nuclear densities for the creation of a still more accurate, dense $\{R,{\bf q}, \omega\}$ surface.

Furthermore, with the Short-Time Approximation supporting identical microscopic numerical simulation structures for \textit{both} electromagnetic and electroweak interactions, multiple model predictions for electron and $\nu$ scattering can eventually be compared to assess overall validity of theory against experiment, allowing for better understanding of modeling systematics and their effects on interpretations of future $\nu$ measurements to take place within the quasielastic regime at future long- and short-baseline $\nu$ oscillation facilities. Such a program, in concert with many actively developing improvements across many simulation types and energy regimes within the broader community, may be able to better elucidate physics beyond the Standard Model. This is especially possible within the Short-Time Approximation formalism due to its extensible nature beyond light nuclei via Auxiliary Diffusion Monte Carlo methods up to \isotope[40]{Ar}. We plan to continue this work beyond \isotope[4]{He} to include \isotope[3]{He}, \isotope[12]{C}, and possibly even \isotope[6]{Li}.

\section{Acknowledgements}

We thank Adi Ashkenazi, Racquel Castillo-Fernandez, Alessandro Lovato, Afroditi Papadapalou, and Gabe Perdue for useful discussions at various stages of this work. We also thank Steve Dytman and Noemi Rocco for clarifications on the scaling of nuclear response functions and how to use them, and Ingo Sick for providing us with the world response data for electron scattering on \isotope[4]{He}. SP has been supported by the U.S.~Department of Energy under contract DE-SC0021027, and through the Neutrino Theory Network (NTN) and the FRIB Theory Alliance award DE-SC0013617. The NTN award was also used to support JLB and SG's visits to Washington University in St. Louis to collaborate on this project. The many-body calculations were performed on the parallel computers of the Laboratory Computing Resource Center, Argonne National Laboratory, the computers of the Argonne Leadership Computing Facility (ALCF) via the 2019/2020 ALCC grant ``Low energy neutrino-nucleus interactions'' for the project NNInteractions, and on resources provided by the Los Alamos National
Laboratory Institutional Computing Program.  This document was prepared by members of the {\fontfamily{qcr}\selectfont GENIE} QMC STA implementation working group using the resources of the Fermi National Accelerator Laboratory (Fermilab), a U.S. Department of Energy, Office of Science, HEP User Facility. Fermilab is managed by Fermi Research Alliance, LLC (FRA), acting under Contract No. DE-AC02-07CH11359. JLB's work to begin this project was fully supported by the U.S. Department of Energy, Office of Science, Office of Workforce Development for Teachers and Scientists, Office of Science Graduate Student Research (SCGSR) 2019 program. The SCGSR program is administered by the Oak Ridge Institute for Science and Education for the DOE under contract number DE‐SC0014664. JLB is grateful to the US DOE SCGSR Program and Fermilab for their generous support of this and other work. JLB's finishing work on this project was partially supported through 2020 by the Visiting Scholars Award Program of the Universities Research Association. Any opinions, findings, and conclusions or recommendations expressed in this material are those of the authors and do not necessarily reflect the views of the Universities Research Association, Inc. JLB would also like to thank the University of Tennessee's Faculty Senate Research Council for their partial support of this work through the 2020 Summer Graduate Research Assistantship.

\bibliography{bibliography}

\clearpage
\onecolumngrid
\setlength{\parindent}{0pt}
\section{Appendices}
\label{sec:Appendix}
Here, we will expound more technically on elements of our interpolation techniques, including the scaling and alignment behavior of our given nuclear response functions \citep{Pastore:2019urn}. Studies of these properties in the QMC STA response functions can be completed utilizing the nonrelativistic formalisms within \citep{Rocco:2018tes,Rocco:2018vbf,Rocco:2017hmh,Lovato:2020kba,Donnelly_1999,Carlson:2001mp}; this is appropriate, as relativistic effects on such responses (such as a broadening of the QE response distribution) have been shown previously to be rather small \citep{Rocco:2018tes} in many QE-like kinematic regimes, and because the STA itself is currently conceived within a nonrelativistic framework.

\subsection{Scaling Analysis and Densifying $\{R,|{\bf q}|,\omega\}$-space for Expansive Response Interpolation}
\begin{figure}[h]
    \centering
    \includegraphics[width=0.49\columnwidth]{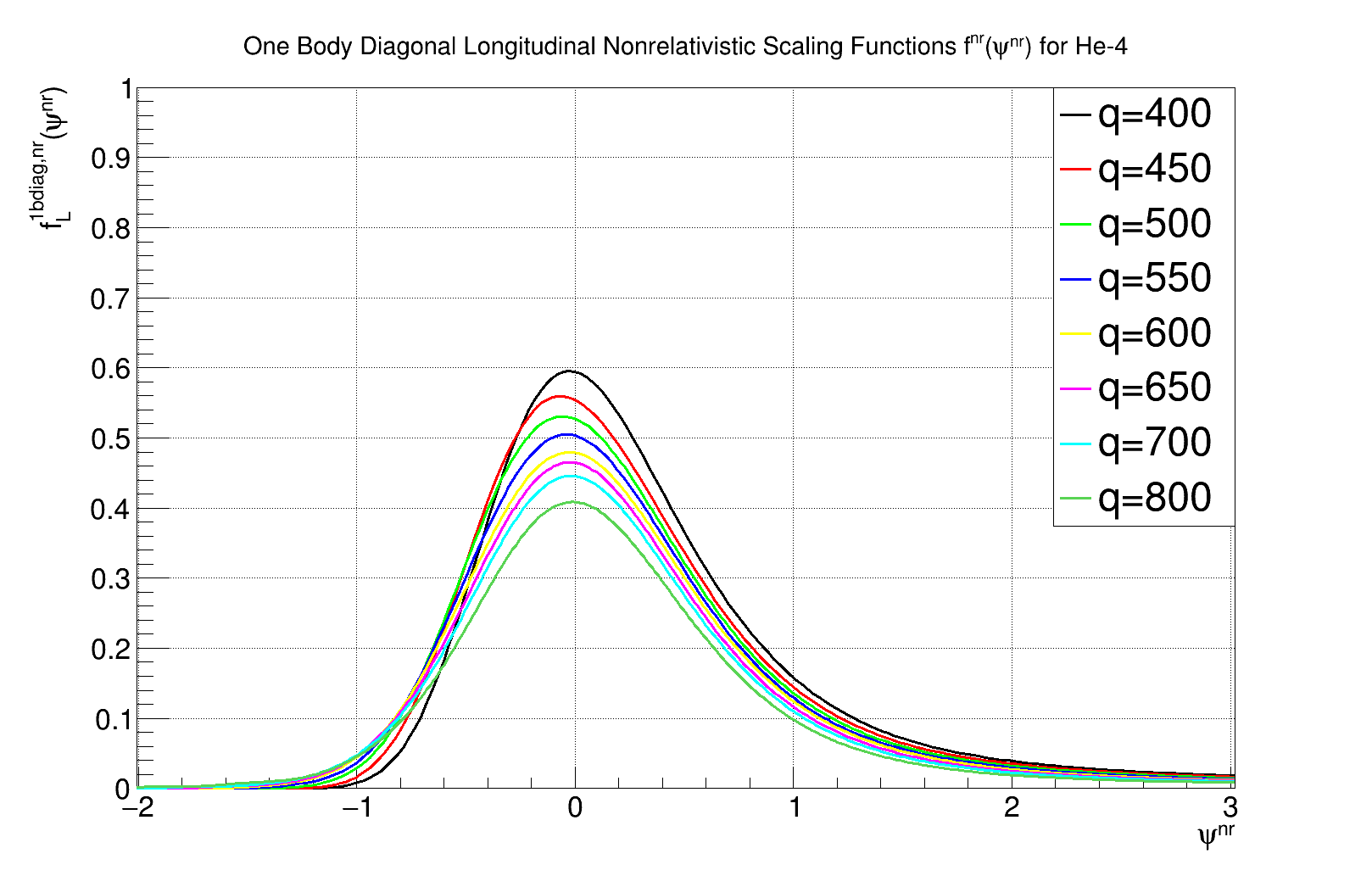}
    \includegraphics[width=0.49\columnwidth]{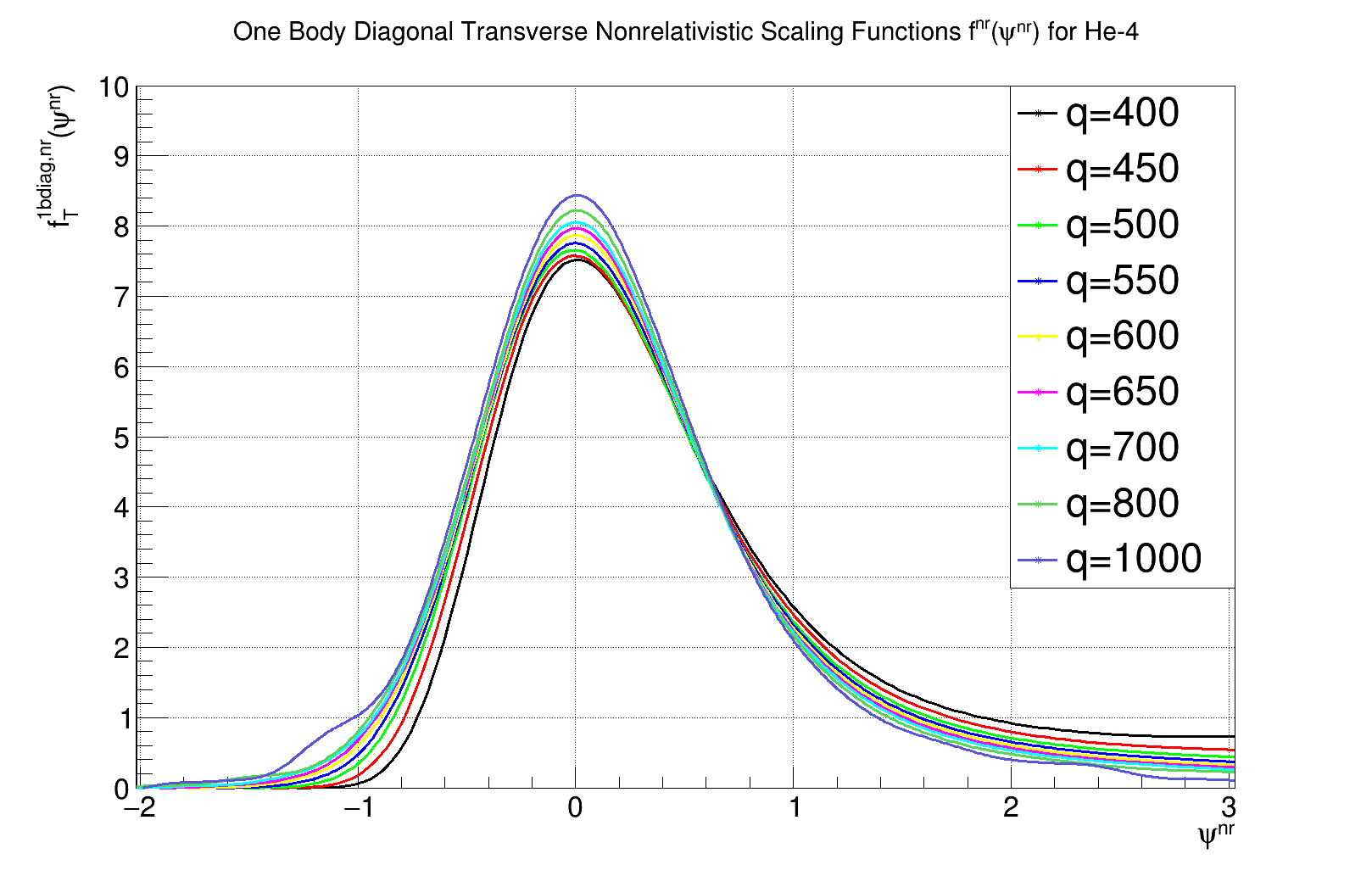}
    \includegraphics[width=0.49\columnwidth]{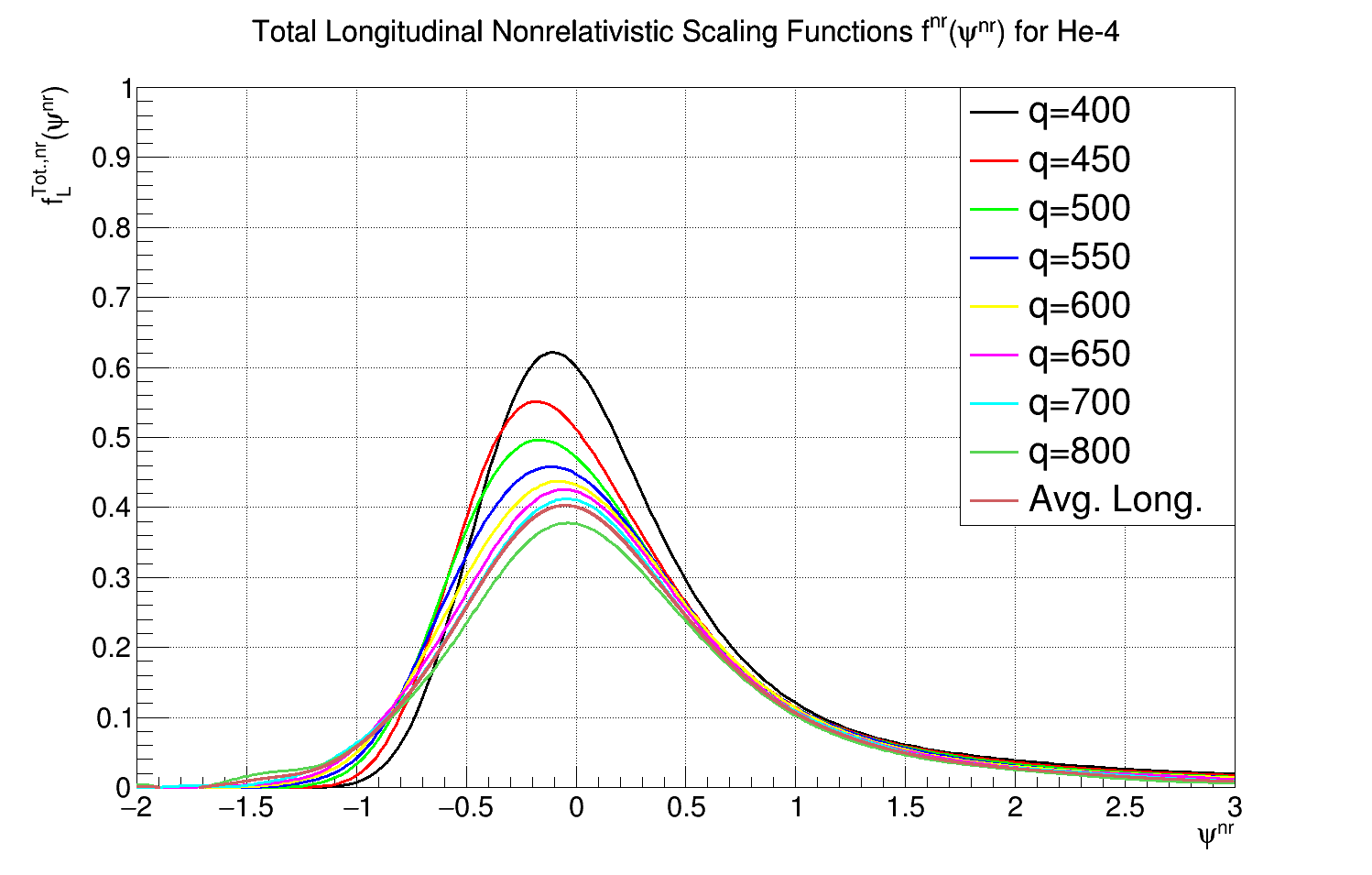}
    \includegraphics[width=0.49\columnwidth]{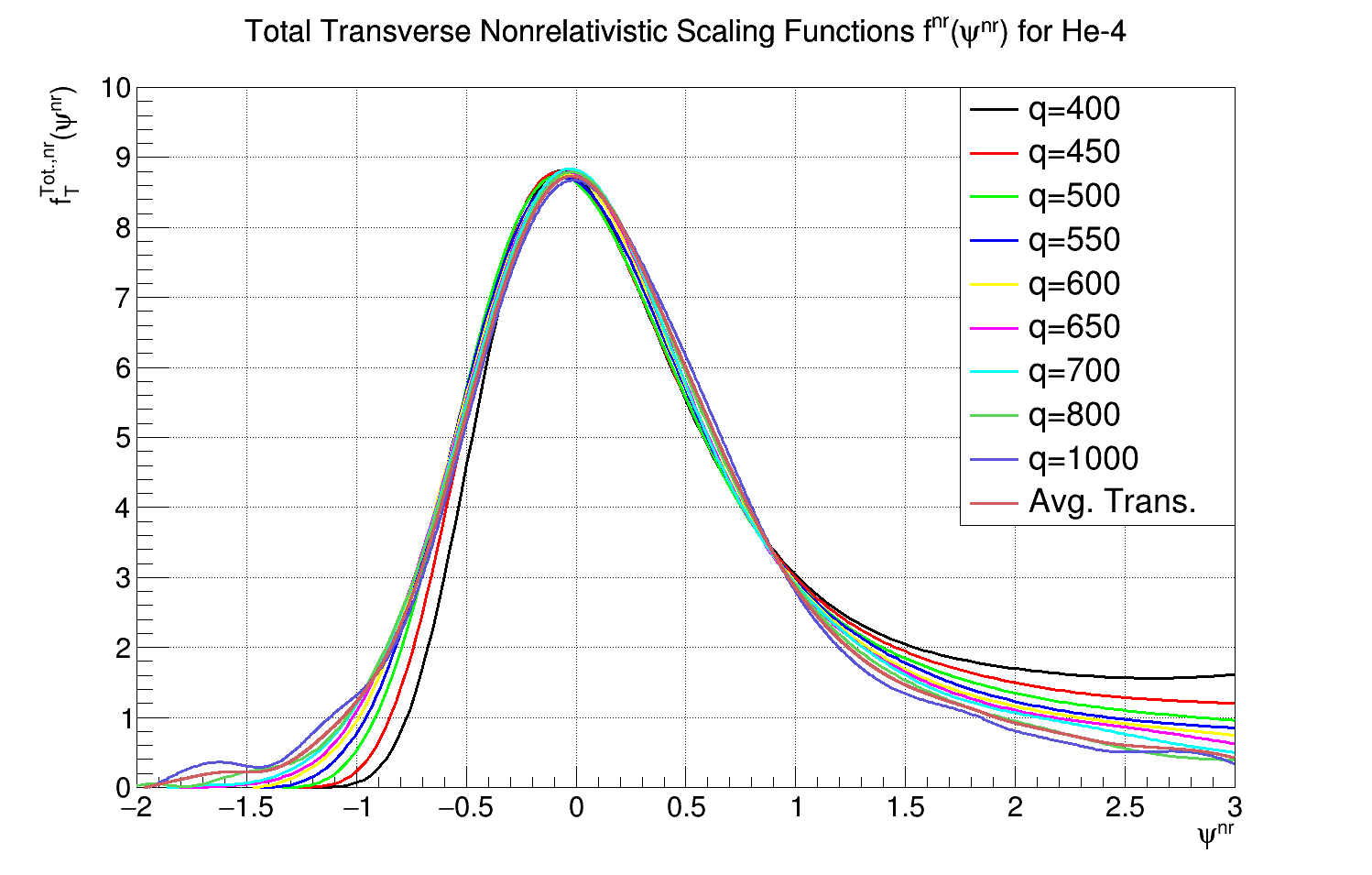}
    \caption{Approximate scaling is observed for both the one-body diagonal term (``1bdiag") and total (``Tot." = one body diagonal + one-body off-diagonal + interference + two-body) electromagnetic response contributions across longitudinal and transverse components; this appears particularly strong in the total transverse response. Note the respective (marginal) destructive and (strongly) constructive behavior of the longitudinal and transverse components when moving from a one-body diagonal to total response paradigm by adding additional interference and two-body terms. The average scaling function is calculated from all shown total responses. Though computed, responses for $|{\bf q}|=\{ 300,350 \}$MeV/c are currently not included in this analysis due to presence of the \textit{elastic} peak, thus spoiling scaling across all components. The longitudinal component of $|{\bf q}|=1000$MeV/c has not yet been computed.}
    \label{fig:1BTotScalingFunctions}
\end{figure}

It is critically important to understand the presence of scaling in the QMC STA response functions given the computational intensity behind the production of even a quite course $\{R,|{\bf q}|,\omega\}$ surface; if present, scaling allows for the (fast, cheap) construction of many finely spaced nuclear response functions, creating a more dense $\{R,|{\bf q}|,\omega\}$ surface which can then be easily interpolated across nearest neighbors to procure any necessary QE kinematic for comparison against empirical QE double differential cross sections. Such is the overarching purpose of the work shown in this appendix, while also serving to confirm expected similarities between the QMC STA and GFMC calculations which utilize the same many-body Hamiltonians, though differing in their computational methods.
\\
\\
The scaling analysis utilizes the {\it nonrelativistic} scaling variable $\psi^{nr}$ \citep{Lovato:2020kba}
\begin{equation}
    \psi^{nr} \equiv \psi^{nr}(|{\bf q}|,\omega)=\frac{m_N}{|{\bf q}| k_F}(\omega-\frac{q^2}{2m_N}-\varepsilon),
    \label{eq:scalingvariablepsi}
\end{equation}
where $m_N$ is the (weighted, nucleus-averaged) nucleon mass, $k_F$ is the approximate Fermi momentum of the system (though this is a somewhat incomplete concept within {\it ab initio} methods), and $\varepsilon$ is included to approximate the binding energy per nucleon of the system (or a corresponding energy shift). It can be conceived that both $k_F$ and $\varepsilon$ may be marginalized over as free parameters, and selected for their optimum scaling behavior; however, for the purposes of this note, we have used values of $k_F=0.18$GeV/c and $\varepsilon=0.015$GeV, as was chosen in \citep{Rocco:2018tes}; more enlightened efforts in the calculation of such constants from known QMC outputs are also possible. Note the particular form of Eq.~(\ref{eq:scalingvariablepsi}) appears in discussions within Ref.~\citep{Lovato:2020kba}, while previous discussions such as those in Ref.~\citep{Rocco:2017hmh} did not take into account this small binding energy shift; removal of this shift does significantly change the scaling behavior.
\\
\\
The actual scaling functions $f_{\alpha}^{nr}(\psi^{nr})$ for a given nonrelativistic response $R_{\alpha}^{nr}(|{\bf q}|,\omega)$ can be considered in the nonrelativistic limit to take the form
\begin{equation}
    f_{\alpha}^{nr}(\psi^{nr}) = k_F \cdot \frac{R_{\alpha}^{nr}(|{\bf q}|,\omega)}{G_{\alpha}^{nr}(|{\bf q}|)}
    \label{eq:scalingfunction}
\end{equation}
where $G_{\alpha}^{nr}(|{\bf q}|)$ can be any component-specific functional combination of single nucleon electric and magnetic form factors \citep{Hohler:1976ax} of neutrons and protons.
From these, and for \textit{symmetric nuclei only} {\it a la} Rocco~{\it et al.}~\citep{Rocco:2017hmh}, we construct the scaling functions when evaluated at the approximate QE peak value of $\omega_{QE}=(\sqrt{q^2+m_N^2}-m_N)$ as
\begin{equation}
    \label{eq:longtransscalingfunctions}
    f_{L}^{nr}=\frac{k_F |{\bf q}| (Q^2+4m_N^2)}{4\mathcal{N}m_N^3} \cdot \frac{R_{L}^{nr}(|{\bf q}|,\omega)}{(G_{E,p}+G_{E,n})^{2}},~~~~
    f_{T}^{nr}=\frac{2 k_F m_N |{\bf q}|}{\mathcal{N}} \frac{R_{T}(|{\bf q}|,\omega)}{q^2 \cdot (G_{M,p}+G_{M,n})^2 + k_F^2 (G_{E,p}+G_{E,n})^2 (1-\psi^{nr})},
\end{equation}
where $\mathcal{N}$ is the number of neutrons or protons in the symmetric nucleus.
Note the possibility of singularities within the transverse scaling function given higher values of $\psi^{nr}\gtrsim4.5$, or $|{\bf q}|\gtrsim400$.
As seen in Figs. \ref{fig:1BTotScalingFunctions}, these scaling formulations appear to (approximately) hold for \textit{both} one-body diagonal and total response contributions across longitudinal and transverse components, showing many similarities to scaling analyses pursued within \citep{Rocco:2017hmh,Lovato:2020kba} over a finite range of $\psi^{nr}$.

\begin{figure}[h!]
    \centering
    \includegraphics[width=0.49\columnwidth]{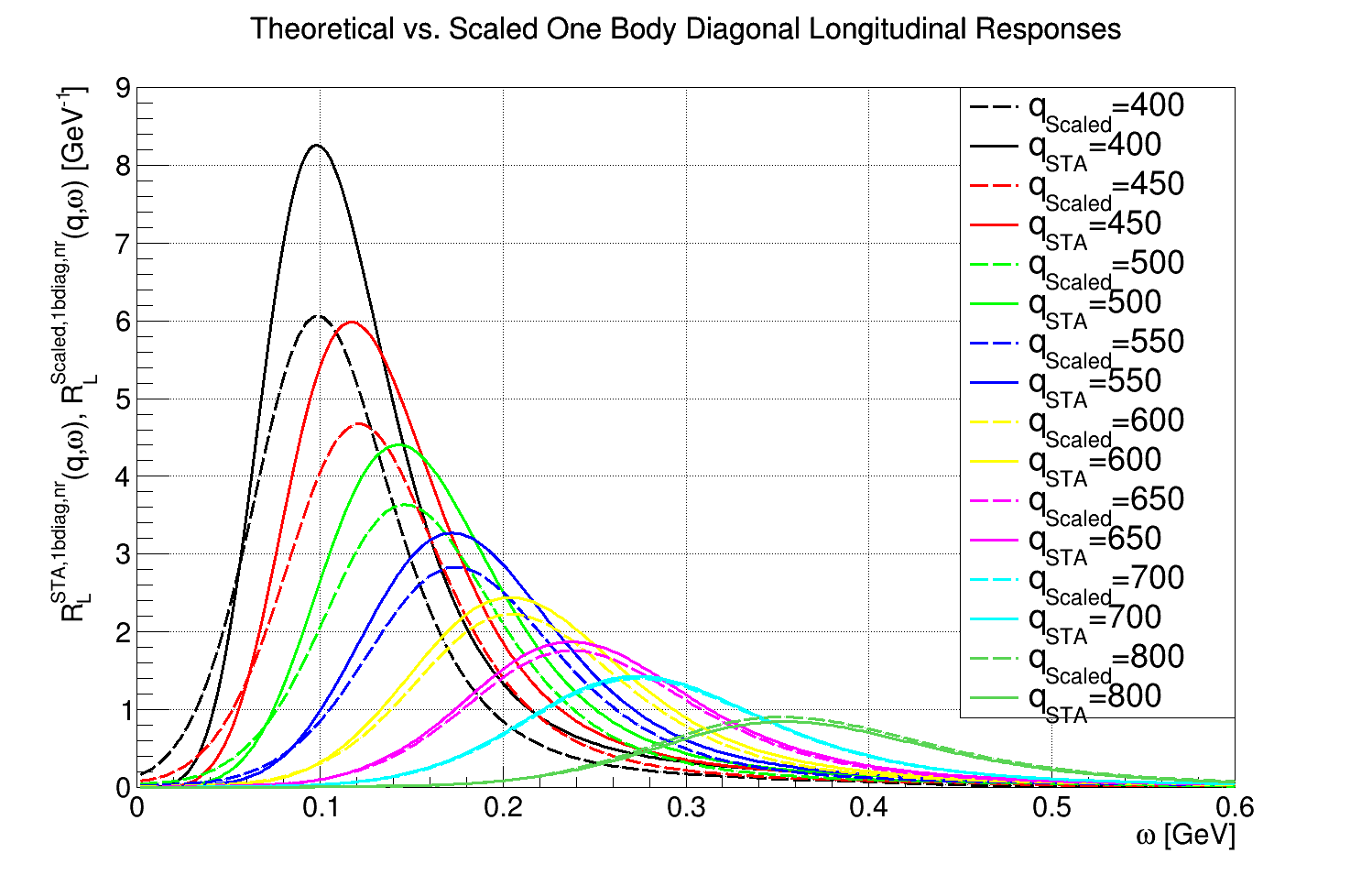}
    \includegraphics[width=0.49\columnwidth]{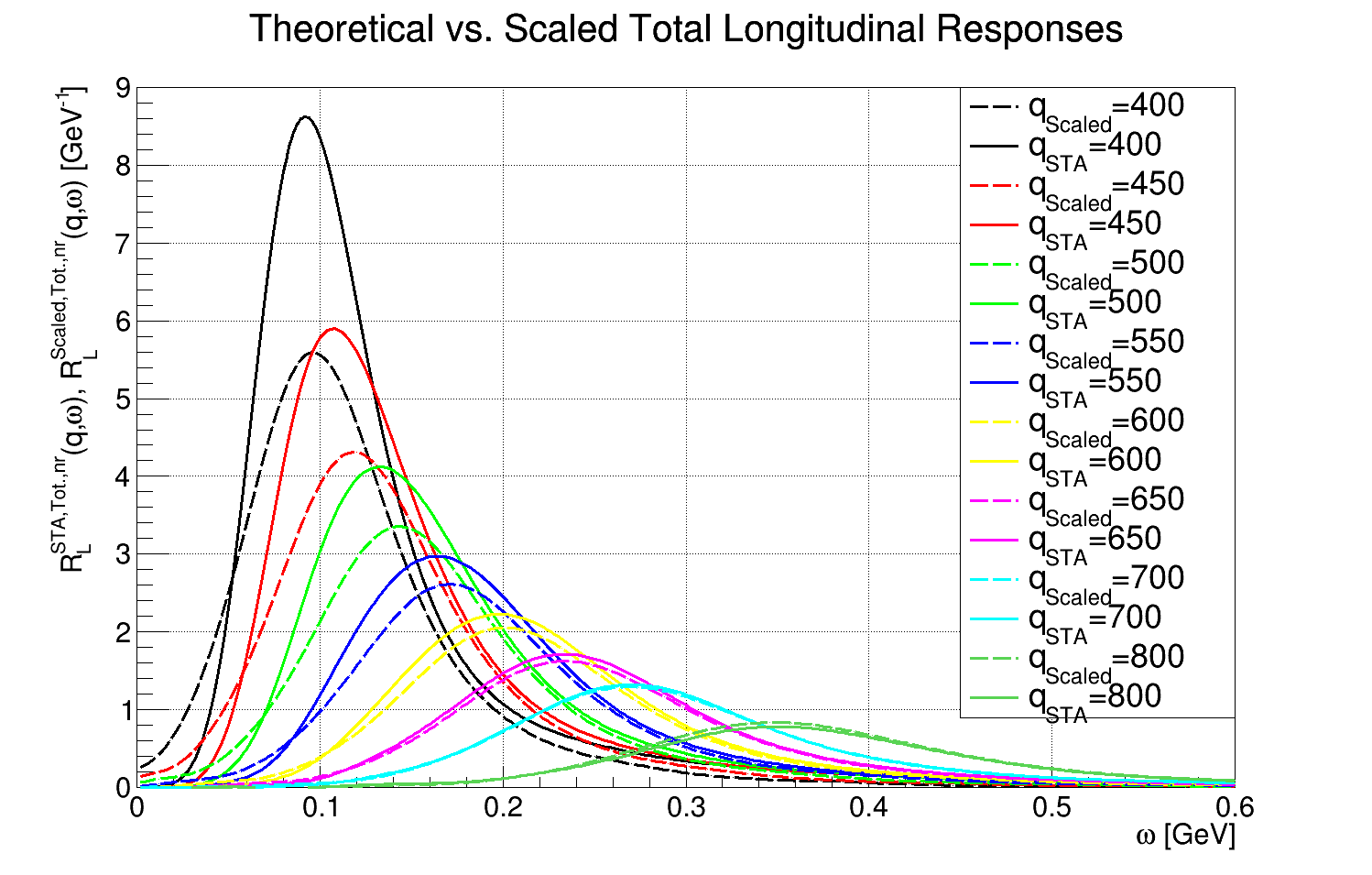}
    \includegraphics[width=0.49\columnwidth]{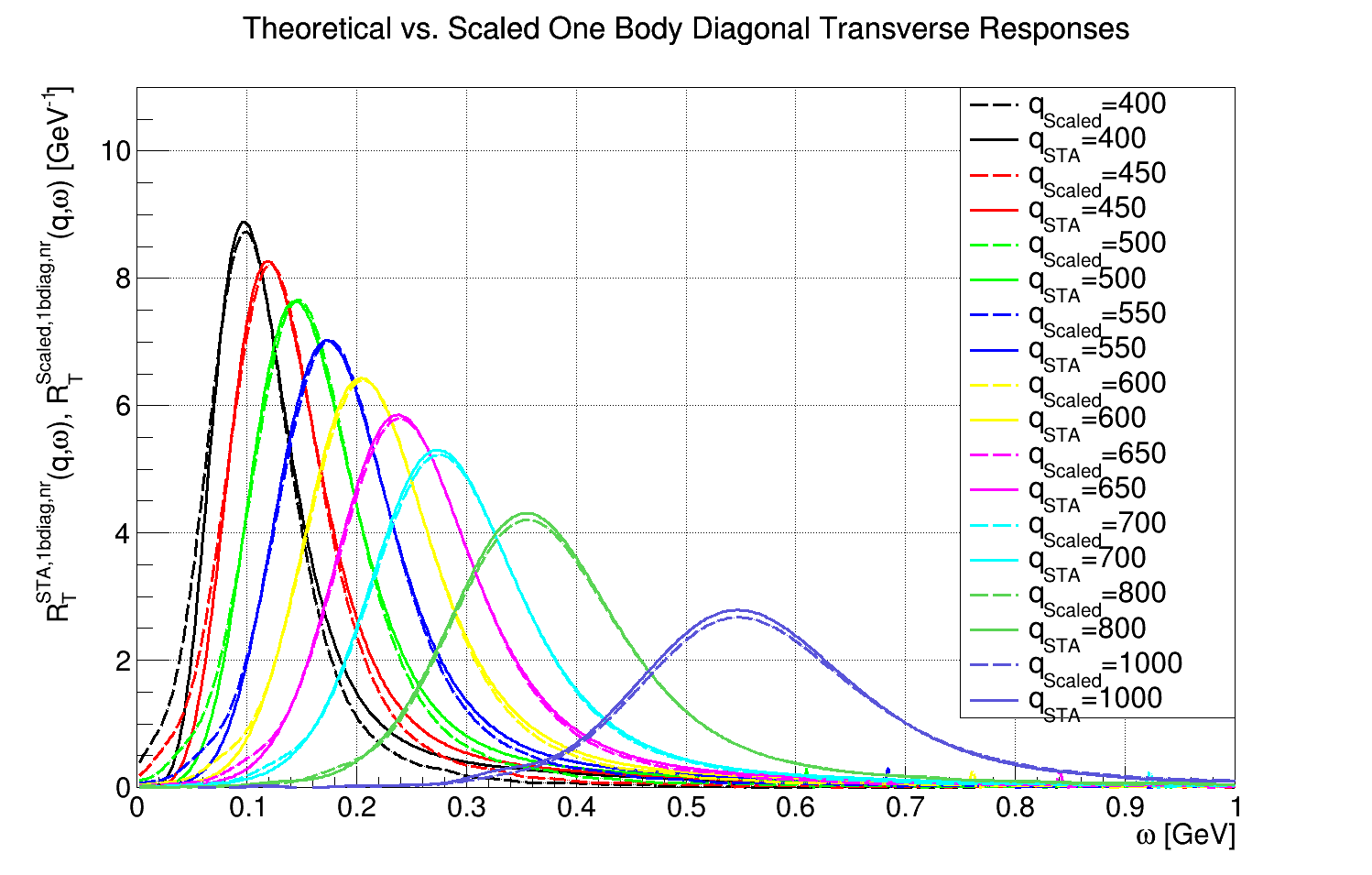}
    \includegraphics[width=0.49\columnwidth]{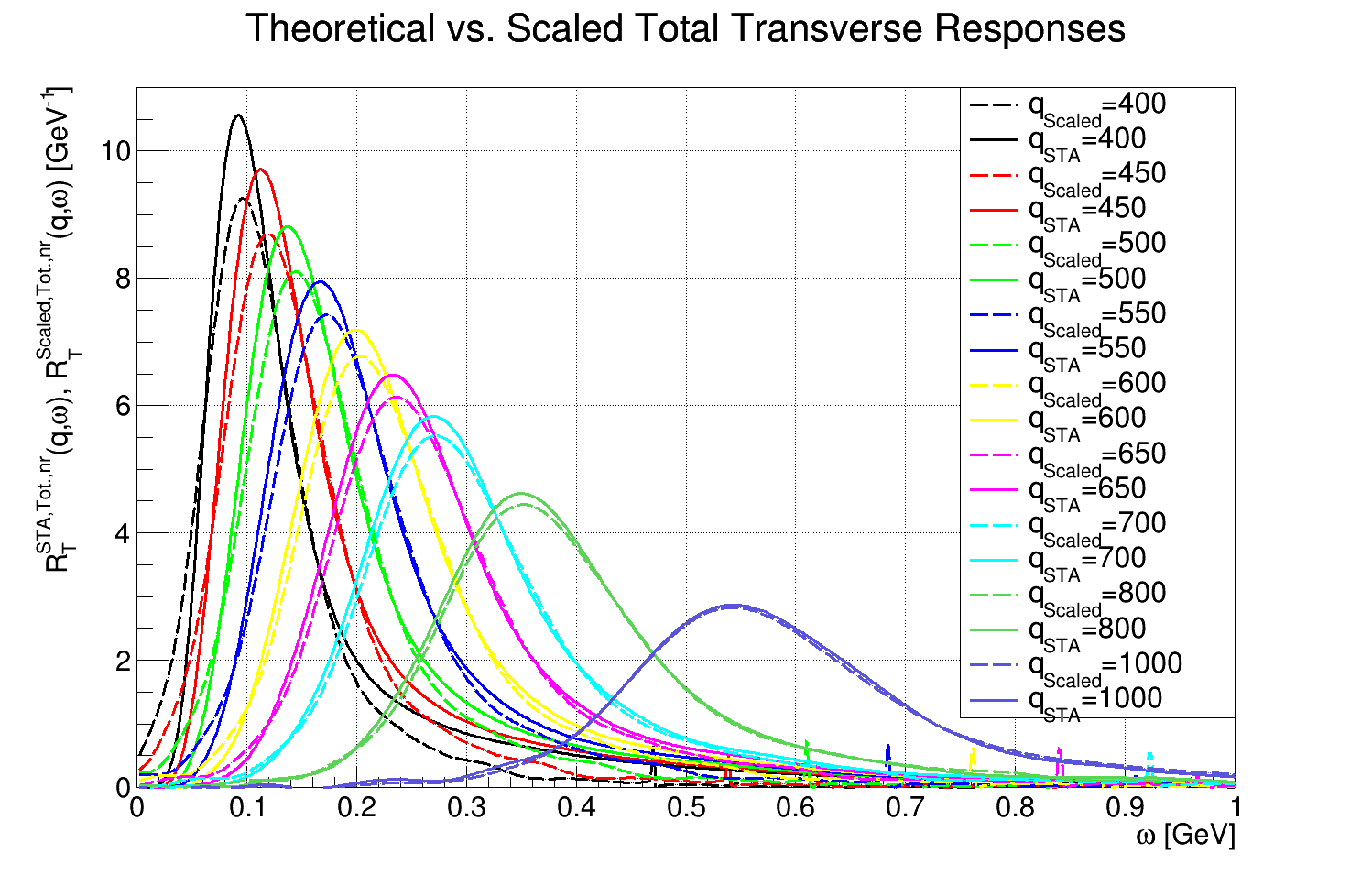}
    \caption{Comparisons between newly created scaled (dashed) and originally computed (solid) one-body diagonal and total nuclear response functions are shown.
    Note the excellent agreement of the transverse responses due to the lack of strength of the elastic peak in this particular component, while the reduced strength of the scaled longitudinal responses removes the elastic strength due to higher momentum transfer responses outweighing the average scaling function; however, too much strength is lost here due to the averaging scheme in both the longitudinal and transverse responses. Other methods may be pursued in future work.}
    \label{fig:1BTotNewvsOldResponseFunctions}
\end{figure}

\begin{figure}[h]
    \centering
    \includegraphics[width=0.49\columnwidth]{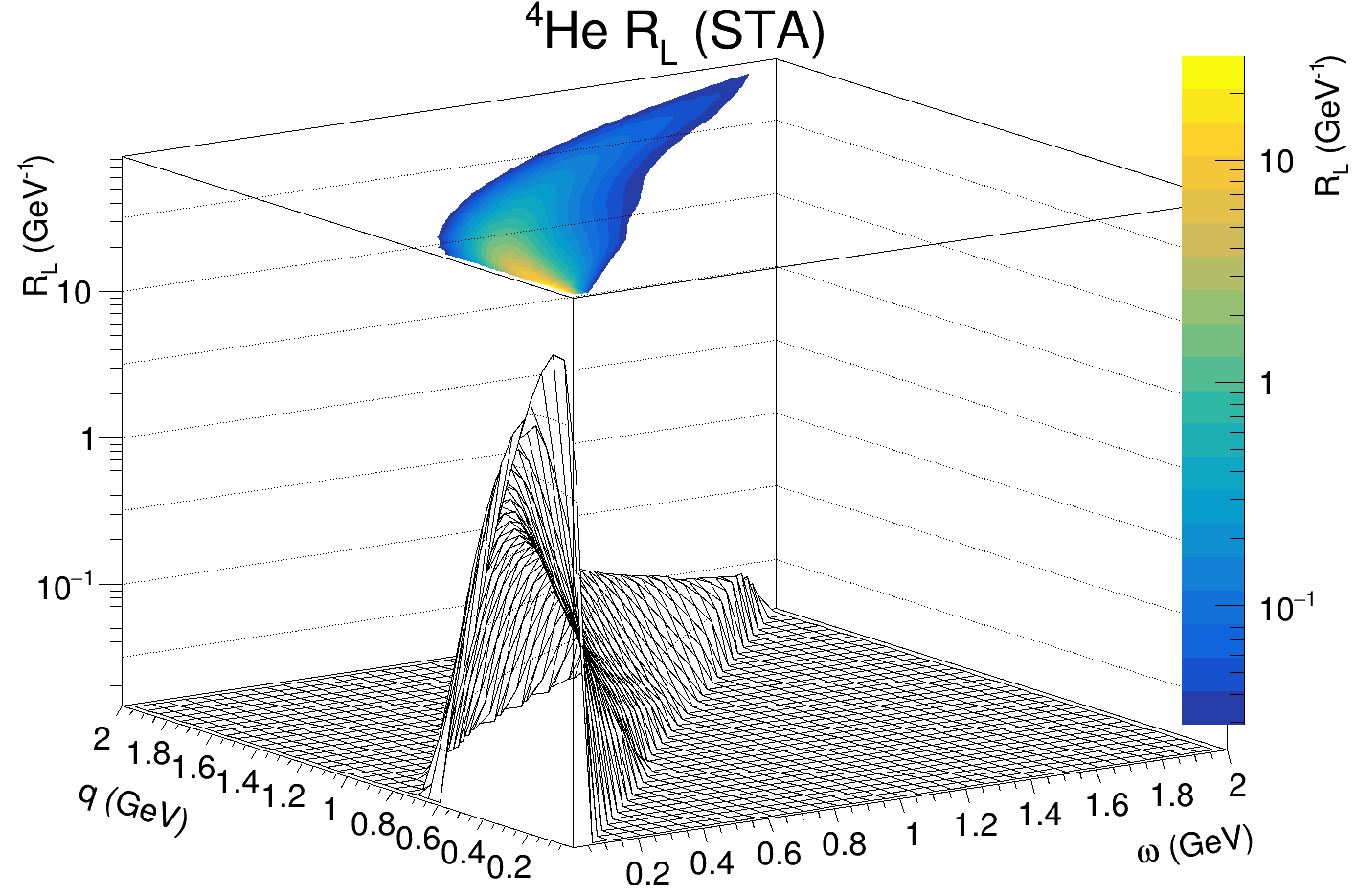}
     \includegraphics[width=0.49\columnwidth]{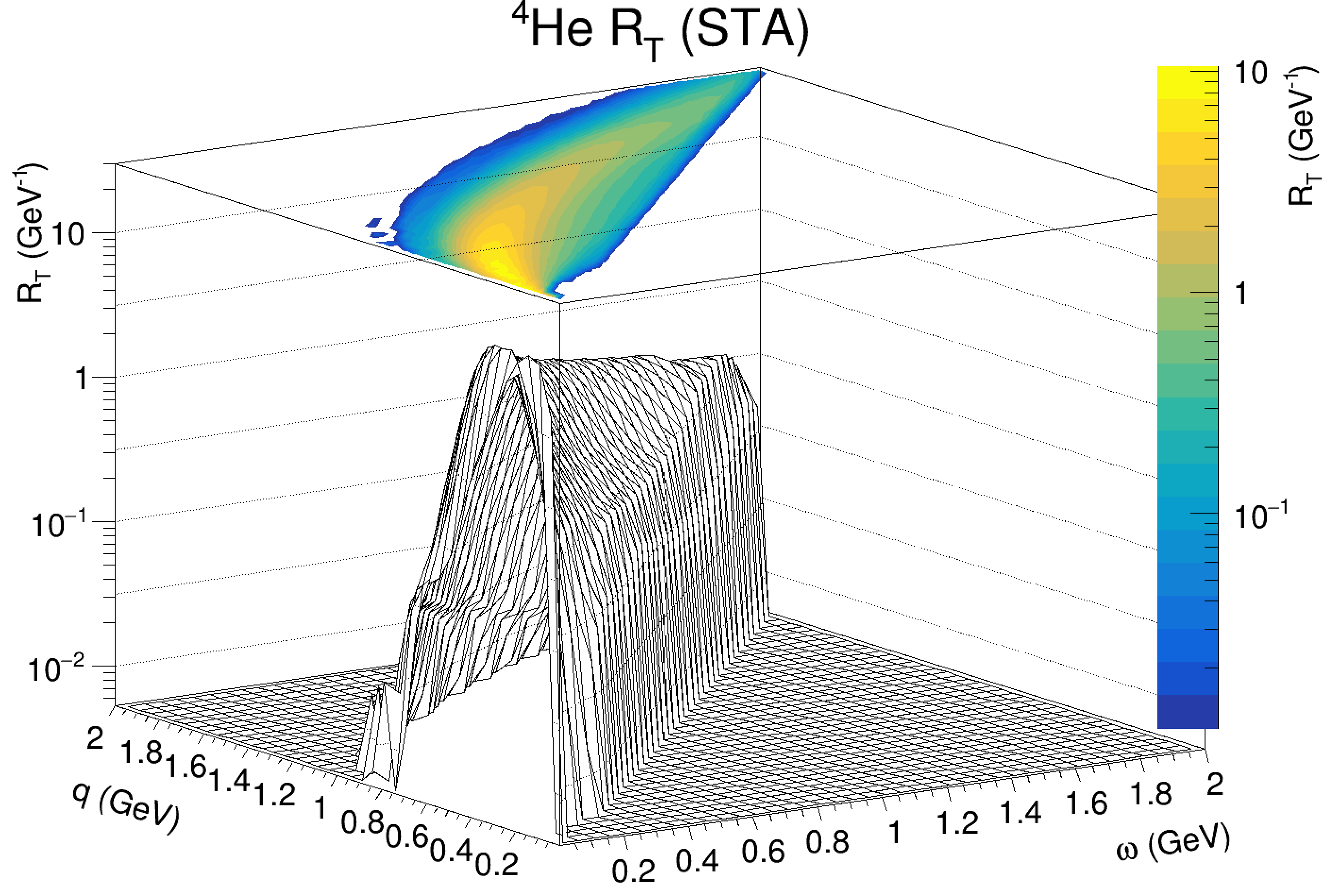}
    \caption{The interpolated nuclear response are shown. Lines along the $\{R,q,\omega\}$ surfaces are visual aides only.}
    \label{fig:ResponseInterpolationAppendix}
\end{figure}

\begin{figure}
    \centering
    \includegraphics[width=0.3\columnwidth]{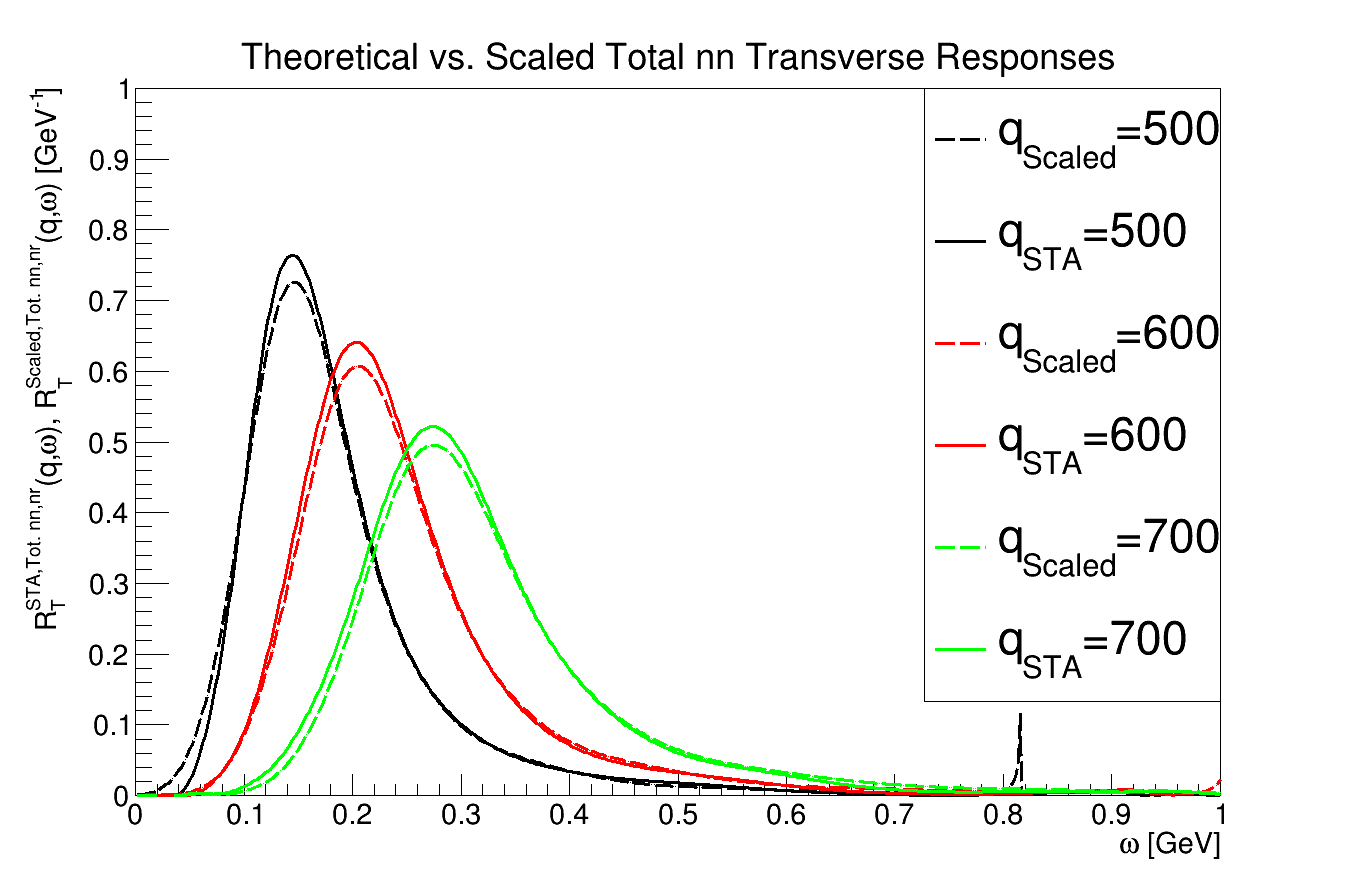}
    \includegraphics[width=0.3\columnwidth]{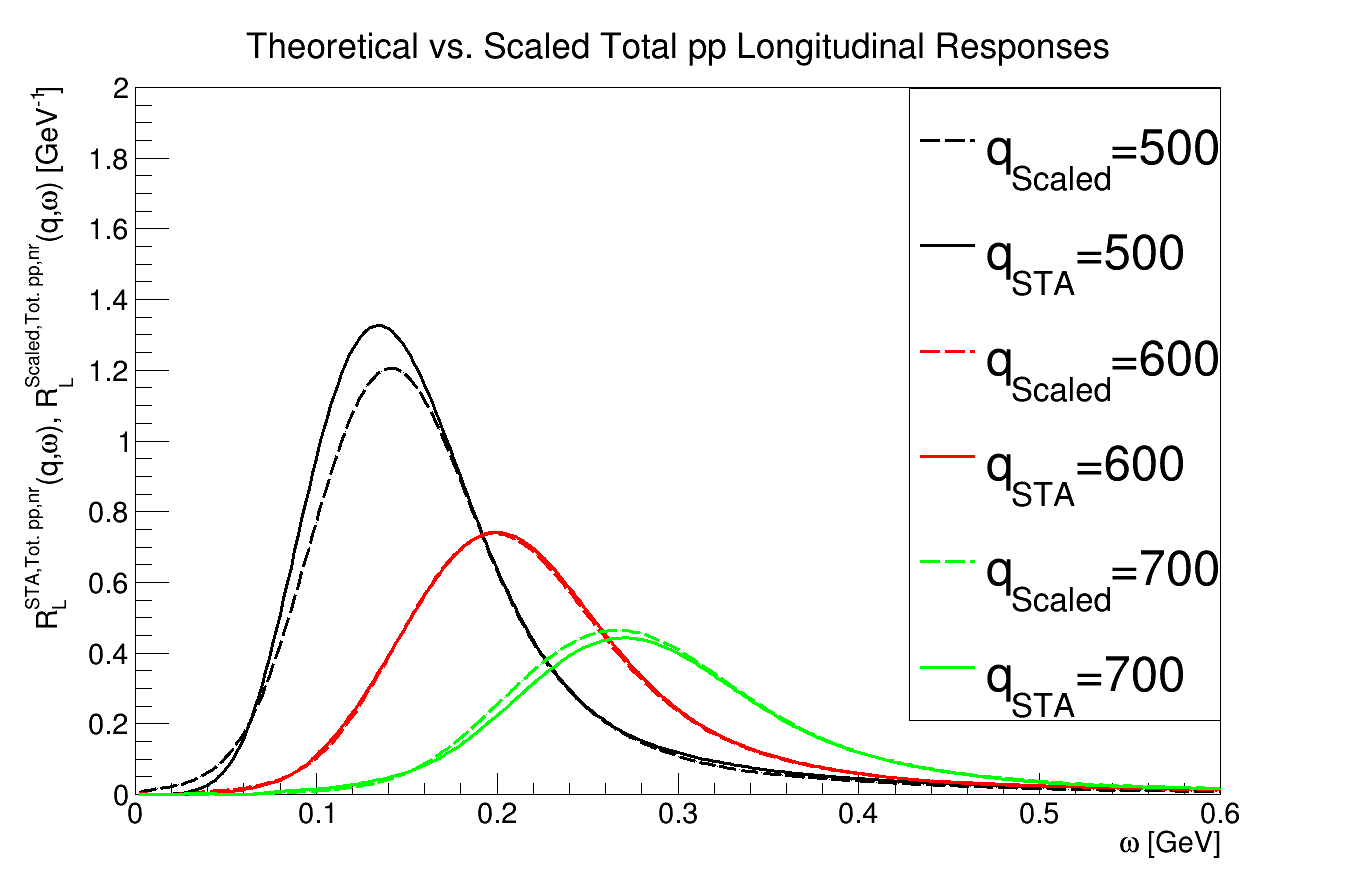}
    \includegraphics[width=0.3\columnwidth]{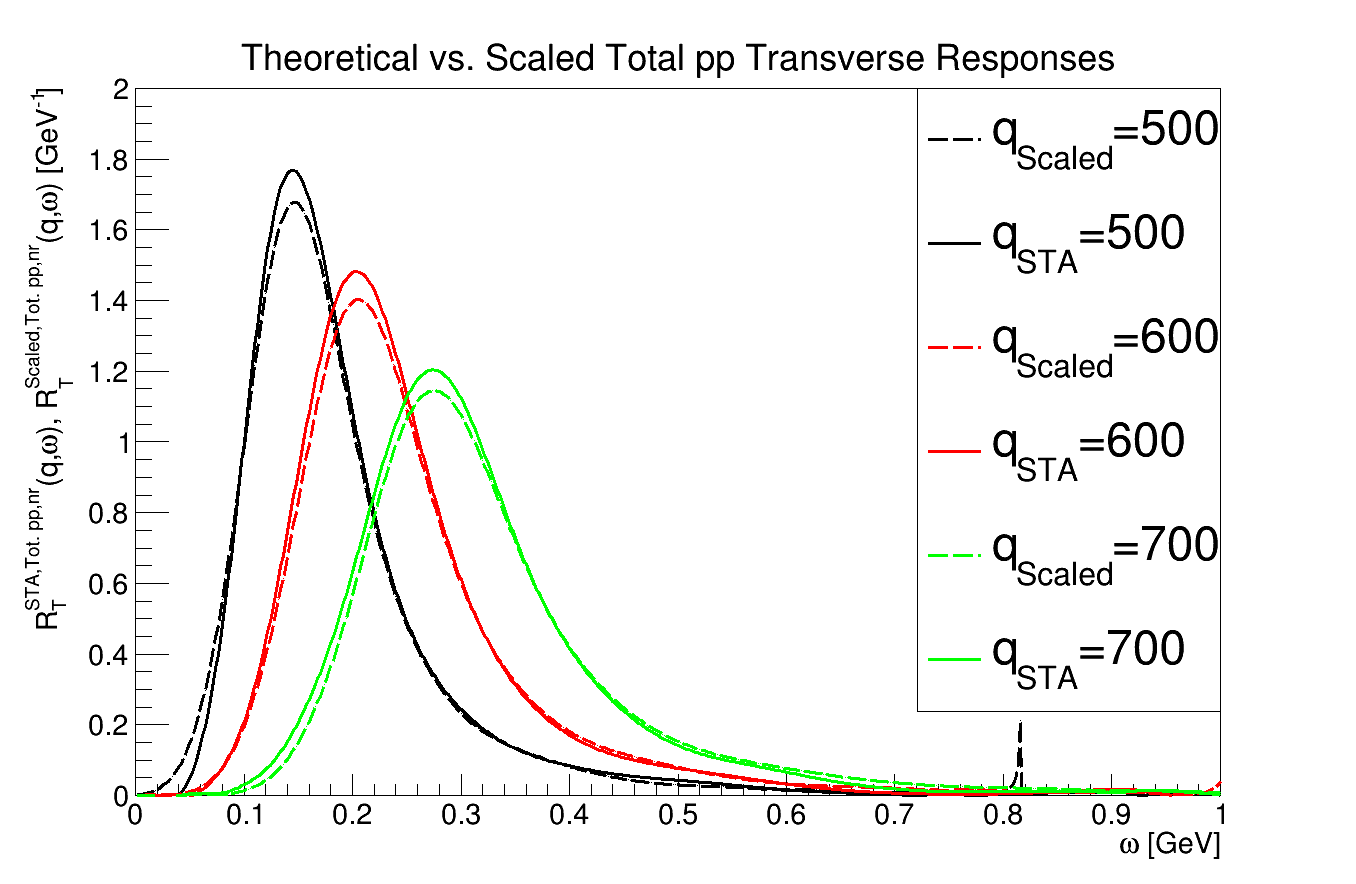}
    \caption{The original \cite{Pastore:2019urn} and scaled two-body particle identity-specific nuclear reponse functions are shown for $pp$ and $nn$ pairs, mirroring Figs. \ref{fig:1BTotNewvsOldResponseFunctions}. We show only $|{\bf q}|=\{ 500,600,700 \}$MeV/c responses here; in principle, the number of known responses can increase, allowing for better-behaved and \textit{expansive} interpolation for a densifying of the two-body $\{ R_{NN}, {|\bf q}|,\omega\}$-surface; from this, more robust double differential cross sections could be derived. Note the different ranges (strengths) of the different components of each channel due to differences in underlying pairing dynamics; the $nn$ longitudinal responses are not shown due to low values.}
    \label{fig:TwoBodyResponses}
\end{figure}

With confirmation of (approximate) scaling behavior ($\sim$invariant shape/alignment across many $|{\bf q}|$-values), one may begin to conceive of an interpolation scheme to create a more dense $\{R,|{\bf q}|,\omega\}$ grid. From a given set of fully computed responses with $|{\bf q}|\in\widetilde{Q}$ of size $N$, a general approximated strategy is to construct an \textit{averaged} nonrelativistic scaling function (visible in total graphs of Figs.~\ref{fig:1BTotScalingFunctions} by burgundy solid lines and labeled as `Avg. Long.' and `Avg. Trans.') as
\begin{equation}
    \overline{f^{nr}_{\alpha}}[\psi^{nr}(|{\bf q}|,\omega)] = \frac{1}{N} \sum^N_{i=1} f_{\alpha,i}[\psi^{nr}(|{\bf q}_{i}|,\omega\in\widetilde{Q})],
    \label{eq:averagedscalingfunction}
\end{equation}
such that one may invert Eq.~(\ref{eq:scalingfunction}) to extrapolate many responses from this single, $|{\bf q}|$-independent, averaged nonrelativistic scaling function using
\begin{equation}
    R_{\alpha}^{nr}(|{\bf q}|,\omega)=\frac{1}{k_F}\cdot G_{\alpha}(|{\bf q}|) \cdot \overline{f^{nr}}[\psi^{nr}(|{\bf q}|,\omega)].
\end{equation}
The average scaling functions $\overline{f^{nr}_{\alpha}}(\psi^{nr})$ are calculated from individual components of the total response scaling functions by a simple unweighted average. Weighted averaging, or effectively choosing \textit{which} scaling functions are best behaved, has not yet been investigated, though in principle could be done so to by marginalizing over some parameter(s) in the average's coefficients and comparing against experimental data in an automated way. The presence of scaling, or effective $|{\bf q}|$-invariances, permits an \textit{expansive} formulation of new $R_{\alpha}^{nr}(|{\bf q}|,\omega \notin \widetilde{Q})$, and allows one to interpolate \textit{between and beyond} the limited known response values at particular $|{\bf q}|\in\widetilde{Q}$, a critical component for QE event generation at many kinematics within {\fontfamily{qcr}\selectfont GENIE}.
\\
\\
Using this method, one can compare the \textit{original} computed nuclear response functions \citep{Pastore:2019urn} with those outputted from the average scaling function approach, as seen in Figs. \ref{fig:1BTotNewvsOldResponseFunctions} for $|{\bf q}|\in\widetilde{Q}=\{400,450,\ldots,750,800,1000\}\,$MeV. Overall agreement of the one-body diagonal longitudinal and transverse components are quite good, as expected from the studies of Ref.~\citep{Rocco:2017hmh}, especially at progressively higher momentum transfers where scaling behavior is maximized \citep{Donnelly_1999} and the presence of the elastic peak in the longitudinal response is absent.
Given the necessity of filling out the $\{R,|{\bf q}|,\omega\}$ surface (especially at higher $|{\bf q}|$-values) for more accurate active nearest-neighbors bilinear interpolation within {\fontfamily{qcr}\selectfont GENIE} to create double differential QE cross sections, thousands of these new responses are computed and collated to a form tabulated grid with a fine granularity. Here, we choose the characteristic spacing of $\Delta|{\bf q}|=1\,$MeV over $|{\bf q}|\in\{1,2000\}\,$MeV, and the characteristic spacing of $\Delta\omega=2\,$MeV over $\omega\in\{2,1800\}\,$MeV, providing ample information for good predictions and validation against available world QE EM scattering data. The full sequence of all 4000 newly interpolated responses inhabiting the full $\{R,|{\bf q}|,\omega\}$ space from transverse and longitudinal components can be seen in Figs. \ref{fig:ResponseInterpolationAppendix}. These are completed with average scaling input in $1\,$MeV spacing, and GENIE is allowed to bilineraly interpolate these on $0.5\,$MeV intervals (the example ``thrown" energy for the QMC STA QE event generator).
\\
\\
This same method can be repeated on pairs of nucleons with known particle identities, such as $pp$ and $nn$ pairs \cite{Pastore:2019urn}. This is especially possible for these pairs given the relative lack of two-body correlations present between them, allowing scaling behavior to more readily manifest. Average $pp$ and $nn$ scaling functions can be constructed from known $|{\bf q}|=\{ 500,600,700 \}$MeV/c two-body particle identity-specific nuclear response functions. These will in principle grow in accuracy when a more complete set (perhaps $N>7$) of response functions are computed with finer $50\,$MeV spacing. The current method leads to the curves seen in Figs. \ref{fig:QETwoBodyXSec}, but more can bee seen in Figs. \ref{fig:TwoBodyResponses}.

\section{Future Work}
\subsection{Nuclear Responses in $\{R/G_{E,p}^2, |{\bf q}|,\psi^{'} \}$-space for \textit{Aligned} Interceding Response Interpolation}
\begin{figure}
    \centering
    \includegraphics[width=0.49\columnwidth]{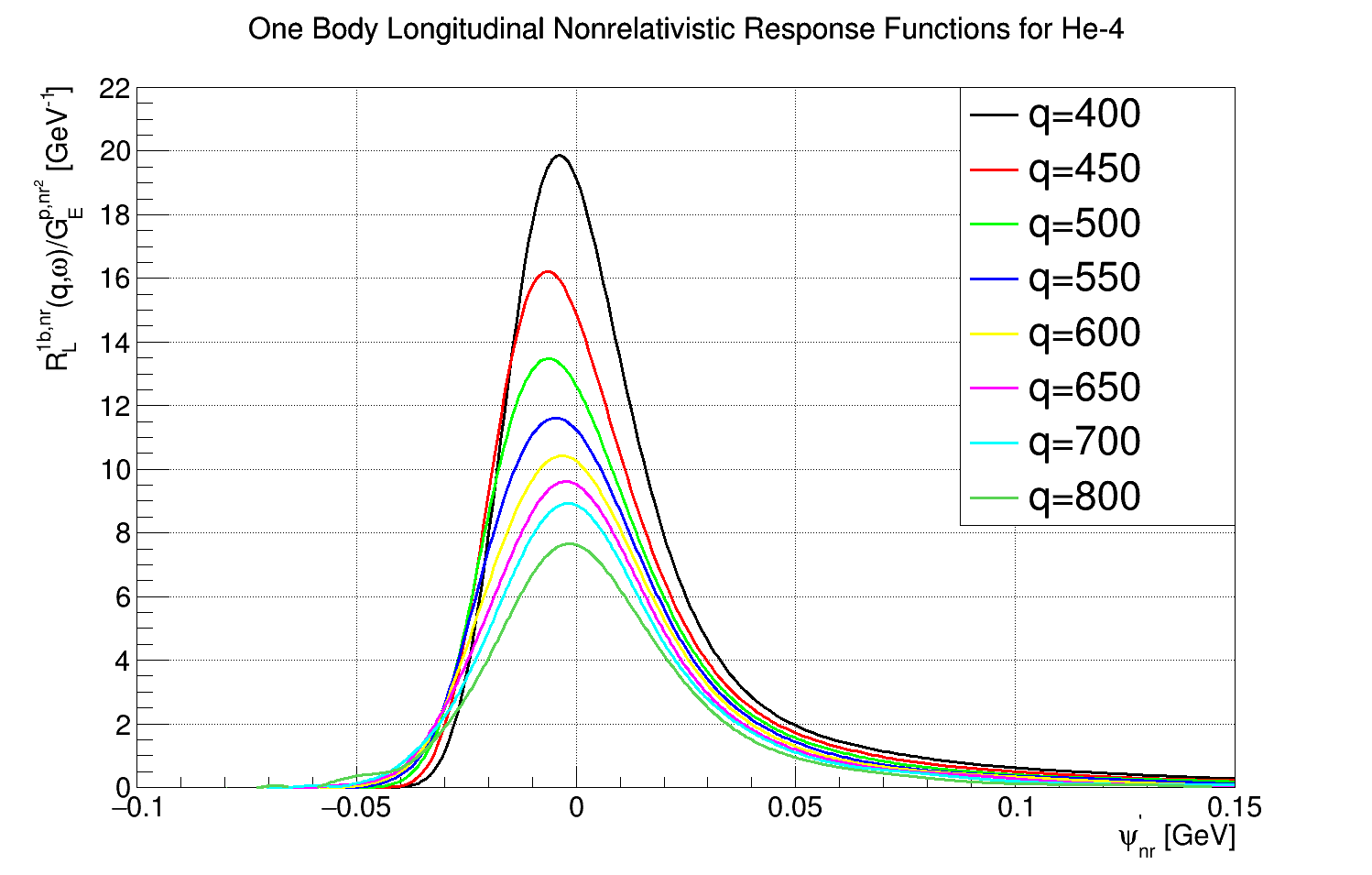}
    \includegraphics[width=0.49\columnwidth]{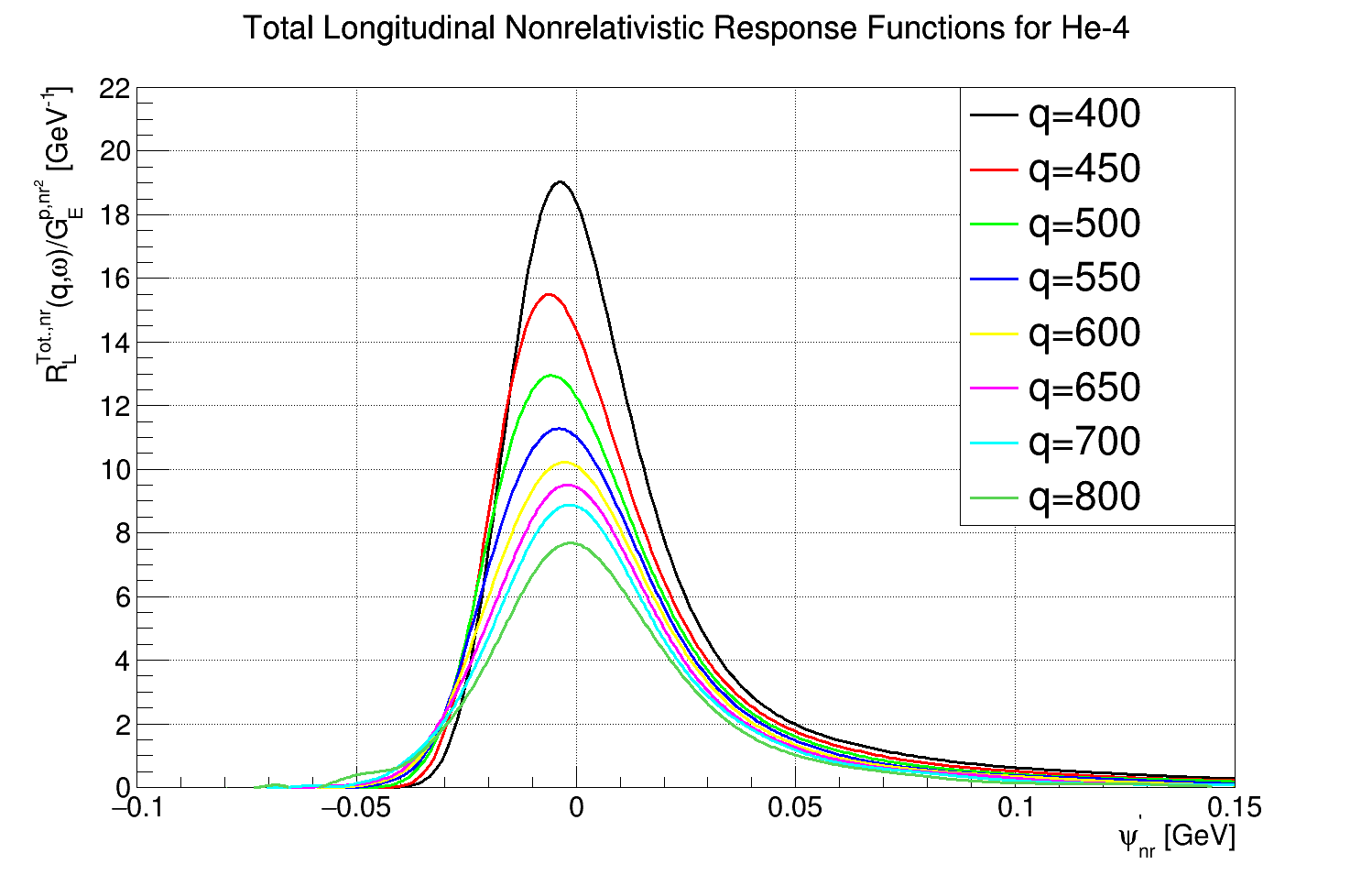}
    \includegraphics[width=0.49\columnwidth]{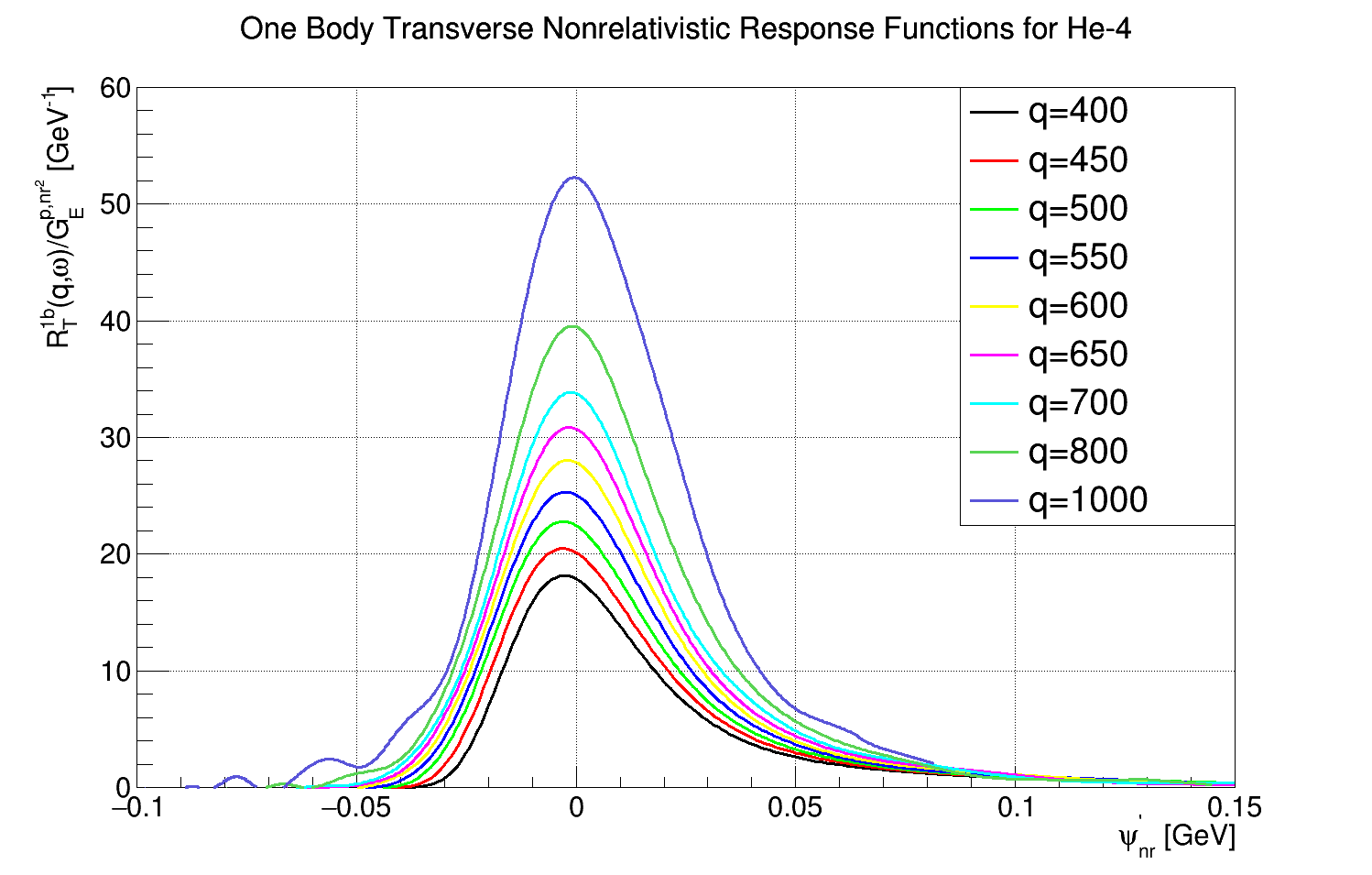}
    \includegraphics[width=0.49\columnwidth]{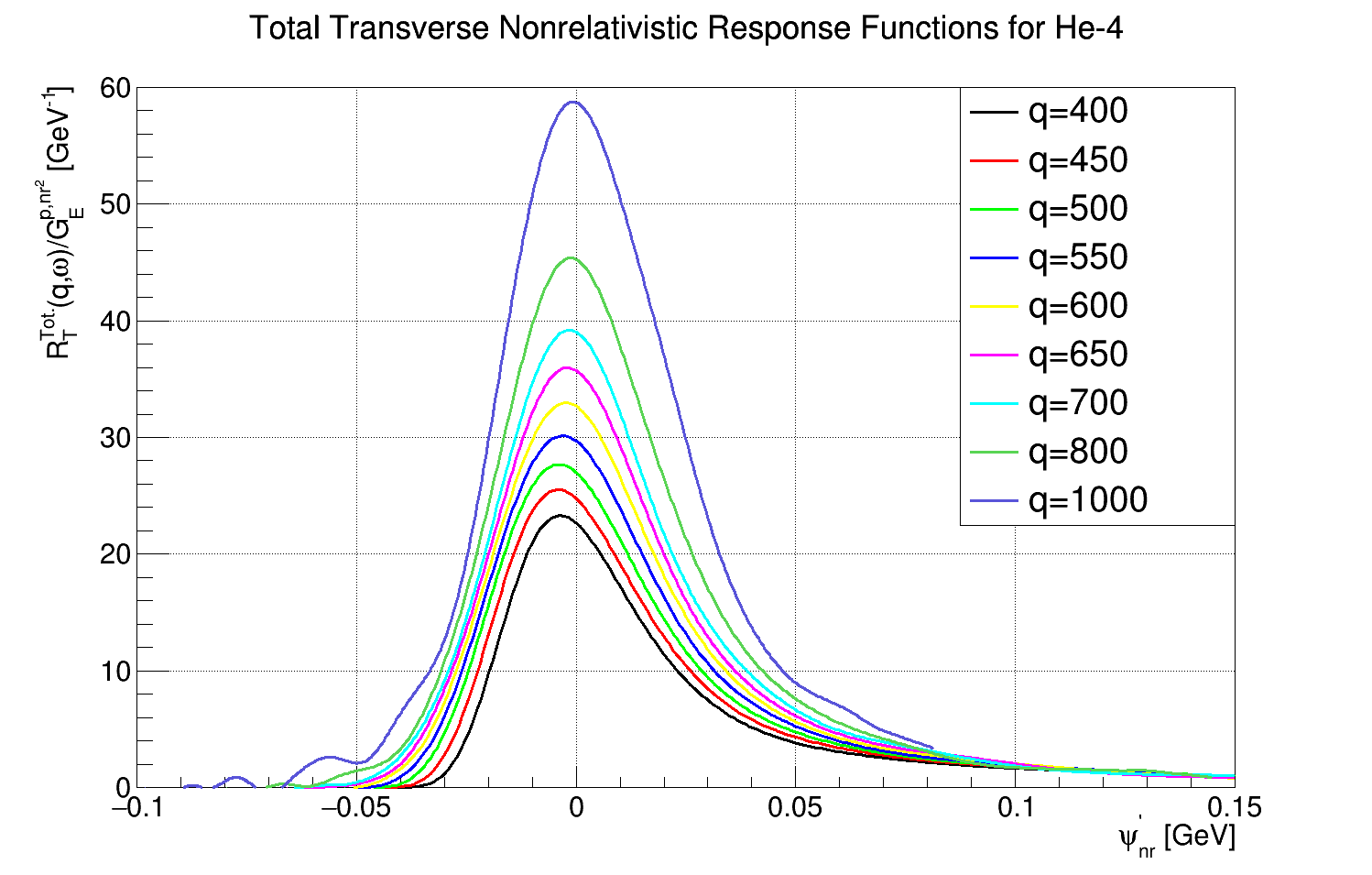}
    \caption{Alignment of form factor normalized response functions is shown when graphing against the nonrelativistic dimensional parameter $\psi^{'}_{nr}$.}
    \label{fig:1bTotResponseAlignment}
\end{figure}

Another method for faster calculation of many responses at many different $|{\bf q}|\notin\widetilde{Q}$ is explored in \citep{Rocco:2018tes} (drawing on previous works \citep{Alberico:1988bv,Barbaro:1998gu,Rocco:2017hmh}), particularly with respect to Eq.~(18) and Figs. 5 and 6 therein; reproductions of these from present work can be seen in Figs. \ref{fig:1bTotResponseAlignment}, where response \textit{alignment} occurs upon the variable transformation $\omega\rightarrow\psi^{'}_{nr}$; here, $\psi^{'}_{nr}$ takes the nonrelativistic \textit{dimensional} form
\begin{equation}
    \psi^{'}_{nr} = k_F  (\frac{\omega-\varepsilon}{|{\bf q}|}-\frac{|{\bf q}|}{2 m_N}),
\end{equation}
and we again choose $k_F=0.18$GeV/c and $\varepsilon=0.015$GeV as in \citep{Rocco:2018tes}. This allows one to transform the dimensional grids $\{R_{nr}, |{\bf q}|,\omega \}\leftrightarrow\{R_{nr}/G_{E,p}^2, |{\bf q}|,\psi^{'}_{nr} \}\equiv [GeV^2]$ for more accurate interpolation \textit{between} aligned responses without loss of generality.
It should be noted that this interceding interpolation scheme will not have the ability to create as much phase space volume as the scaling method outlined above, as it critically does not rely $|{\bf q}|$-invariance to \textit{expand} beyond the known response domain. Work to implement and compare behavior between this \textit{interceding} interpolation and the previously discussed \textit{expansive} interpolation is ongoing, and will be included in a future publication.

\end{document}